\documentclass[11pt, letterpaper]{article}
\usepackage{graphicx} % Required for inserting images
\usepackage{amsmath, amsthm, amssymb}
\usepackage{hyperref}
\usepackage[left=1in, bottom=1in, right=1in, top=1in]{geometry}
\DeclareMathOperator*{\argmax}{arg\,max}

\usepackage{array}
\usepackage[authoryear, square]{natbib}
\usepackage{xcolor}
\usepackage{soul}
\usepackage{booktabs}
\usepackage{multirow}
\usepackage{subcaption}

\usepackage{placeins}   % For \FloatBarrier
\usepackage{makecell}

\usepackage{comment}

\usepackage{setspace}
\usepackage[ruled,vlined]{algorithm2e}

\usepackage{mathtools}
\usepackage{amsfonts}
\usepackage{mathrsfs}

\usepackage{epsfig}
\usepackage{epstopdf}
\usepackage{adjustbox}
\usepackage{enumitem}
\usepackage{authblk}
\usepackage{cleveref}

\theoremstyle{plain}

\newtheoremstyle{uprightremark}
  {}                             % Space above
  {}                             % Space below
  {\normalfont}                  % Body font (upright)
  {}                             % Indent amount
  {\bfseries}                    % Theorem head font
  {.}                            % Punctuation after theorem head
  { }                            % Space after theorem head
  {}                             % Theorem head spec
\theoremstyle{uprightremark}

\theoremstyle{definition}

\begin{document}

\title{Identification of Emotionally Stressful Periods Through \\
Tracking Changes in Statistical Features of mHealth Data}
\author[1]{Younghoon Kim}
\author[2]{Sumanta Basu}
\author[3]{Samprit Banerjee$^\dagger$}

\affil[1]{Department of Mathematics, University of Alabama, Tuscaloosa, AL, USA}
\affil[2]{Department of Statistics and Data Science, Cornell University, Ithaca, NY, USA}
\affil[3]{Department of Population Health Sciences, Weill Cornell Medicine, New York City, NY, USA}

\date{\today}
\maketitle
\def\thefootnote{$\dagger$}\footnotetext{Corresponding author. Email: sab2028@med.cornell.edu}

\maketitle

\begin{abstract}
% Significance:
Identifying periods associated with emotional stress in older patients with mood disorders and chronic pain is important in mental health research. 
% Significance + Rigor of prior research
Existing approaches often rely on conventional change point detection methods, require explicit modeling of relationships between variables, and return discrete change point estimates. These limitations are problematic when distributional shifts are complex, dependencies are difficult to specify, or changes occur asynchronously across series.
% Problem Statement:
We propose a hotspot detection framework, where hotspots are temporal regions in which statistical features of passive sensing variables and stress indicators change at the same or nearby times. 
% Approach
We extend the moving-sum scheme to detect changes within and across series, and construct hotspots using distance-based test statistics and confidence intervals. 
% Innovation
The method does not require a specific functional relationship between series and summarizes changes as intervals rather than isolated time points.
% Validation
Simulation studies show that the proposed Joint- and Bi-MOSUM procedures generally achieve higher power and better control of false detections when multiple types of changes coexist. The two hotspot rules also provide complementary summaries. 
% Result
We further illustrate the method using ALACRITY Phase I data, where the proposed approach identifies complex distributional changes in stress levels that occur in temporal proximity to changes in passive sensing measures.
\end{abstract}


%%%%%%%%%%%%%%%%%%%%%%%%%%%%%%%%%%%%%%%%%%%%%%%%%%%
\clearpage

\section{Introduction}\label{sec:intro}

Emotional stress is the psychological strain that occurs when individuals perceive that demands exceed their coping resources. In older adults, it often co-occurs with mood disorders, and the inability to regulate emotional stress could lead to reduced effectiveness of psychotherapy \cite[e.g.,][]{doraiswamy2002spectrum,murphy2016depressive}. In particular, older patients with mood disorders are more likely to experience reduced quality of life, suicidal behaviors, and elevated all-cause mortality. In this context, identifying the moment when older adults with mood disorders have elevated emotional stress is imperative to intervene at the right moment. Such interventions may help prevent an ensuing cascade of negative outcomes, such as lack of response to treatment or worsening of symptoms. With the development of mobile health (mHealth) technologies, collecting real-time data has become easier, enabling more timely monitoring of emotional stress \cite[e.g.,][]{lee2021use,solomonov2025precision}. Digital interventions can also be delivered through mHealth technology with the potential of preventing negative outcomes following a period of high emotional stress. Patients themselves can report their stress levels and related feelings through mobile devices such as smartphones and wearables, and their emotional status in daily life can be tracked without the need for in-person visits to clinical experts. Automatically detecting the onset of emotional stress could also enable timely interventions, allowing clinicians to respond proactively and potentially prevent a marked decline in mood or health.

Passive sensing variables are datasets collected by mHealth devices through sensors without requiring active input from the user. Typical examples include heart rate variability and blood pressure, which can reflect changes in stress levels \cite[e.g.,][]{maharjan2021passive,sheikh2021wearable}. In this study, behavioral indicators are used instead, which can also serve as passive sensing variables. Patterns of physical activity, including body movement, time spent in social activity, and sleep quality, provide complementary information about a person’s state of emotional stress. Behavioral measures are easy to track with widely available commercial devices, do not require research-grade equipment, and can be collected naturally in daily life. By using these passive sensing variables, it is possible to monitor changes in stress levels in real time, enabling timely identification and intervention \cite[e.g.,][]{sano2018identifying,saylam2023quantifying}.

The goal of studying passive sensing variables in this context is thus to uncover their associations with stress, which can later be used for prediction through machine learning (ML) algorithms when self-reported emotional feelings are unavailable. However, it is quite common that directly applying conventional ML methods to these datasets often leads to poor performance \cite[e.g.,][]{vos2023generalizable,bustamante2024precision,del2025analysing}. This is because both passive sensing variables and stress levels are typically non-stationary; their mean and variability may change across different time periods, and these local variations can become obscured when analyzed over the entire observations.

To address this challenge, change point detection \cite[CPD; e.g.,][for comprehensive surveys]{aminikhanghahi2017survey,truong2020selective,cho2022two} can be applied as a pre-processing step to identify time points where data characteristics shift, allowing each segment to be analyzed or labeled separately before being used in ML models. In the context of stress detection from passive sensing variables, pre-defined rule-based methods were widely accepted \cite[e.g.,][]{kyriakou2019detecting,moser2023individual}. 

More recently, CPD algorithms have emerged as a data-driven alternative for identifying transitions in stress levels. For example, \cite{hosseini2022multimodal} applied a CPD algorithm \citep{truong2018ruptures} to their multimodal sensor data to identify abrupt physiological or behavioral changes associated with transitions between low- and high-stress states in nurses during their work shifts. For older patients, \cite{kim2025co} developed a framework combining the CPD algorithm by \cite{matteson2014nonparametric} with a linear model linking passive sensing variables to stress levels, with the goal of improving the prediction of future stress states.

However, existing algorithms still rely heavily on conventional CPD methods and often impose a specific functional relationship between passive sensing variables and stress levels. In contrast, this study focuses on identifying intervals of potentially asynchronous distributional changes across the two series without explicitly modeling their relationship. This distinction is important when detected changes do not occur at exactly the same time, when multiple distributional features change simultaneously, or when the underlying association between the two series is difficult to specify. Missing the exact or nearby timing of such changes can undermine the usefulness of passive sensing variables and pose challenges for their validation as digital biomarkers in future studies \cite[e.g.,][]{mohr2017personal,gomes2023survey}.

\begin{figure}[h]
\centering
 \includegraphics[width=\textwidth,height=0.6\textwidth]{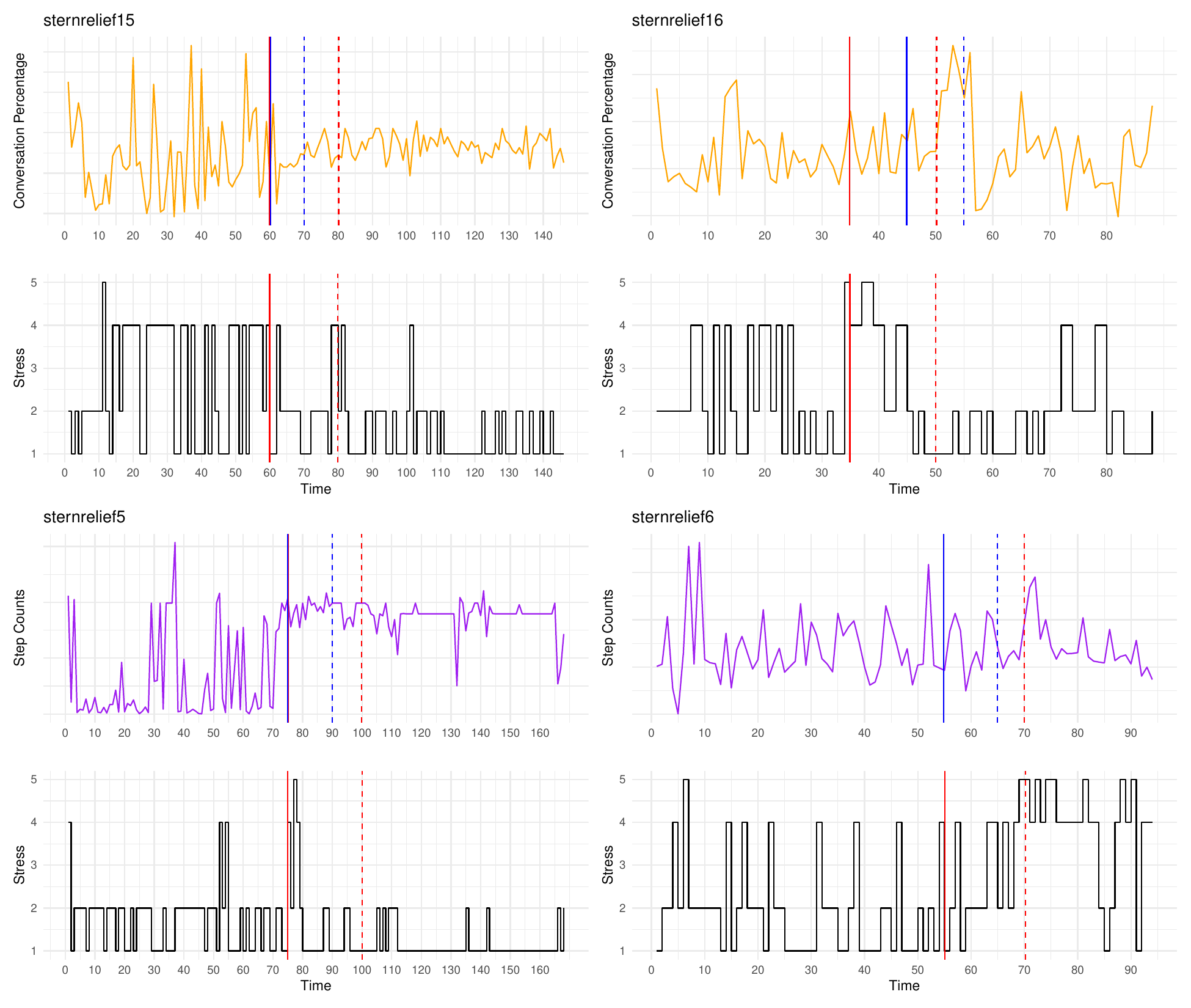}
 \caption{Representative patterns in passive sensing variables and stress levels from four patients in ALACRITY Phase I data. In each panel, the colored lines represent passive sensing variables, and the black lines represent stress levels. The red solid and dashed lines indicate the expected start and end points of changes in stress levels, while the blue solid and dashed lines indicate the expected start and end points of changes in the passive sensing variables.}
\label{fig:patterns}
\end{figure}

\subsection{Motivating Examples}

Examples of pairs of passive sensing variables and stress levels from four different patients are shown in Figure \ref{fig:patterns}. The details of the series are given in Section \ref{sec:data_appl}. The distributional shifts in each of the two series are likely located between the solid and dashed lines, as indicated by visual inspection. However, although the intervals in which these changes occur overlap within each patient’s case, they are asynchronous (i.e., they do not occur at the same time). Moreover, for passive sensing variables, the types of changes in each series cannot be clearly categorized into a single type, such as changes solely in means or variances.

The limitations of conventional CPD algorithms can be summarized as follows. First, traditional CPD methods typically detect only a single type of distributional change (e.g., change in either mean or variance). However, as shown in Figure \ref{fig:patterns}, passive sensing variables and stress levels often exhibit simultaneous changes in both mean and variance. This highlights the need for a more flexible framework capable of identifying local changes across multiple statistical characteristics occurring at the same time. Second, conventional CPD algorithms usually report specific change points, assuming that changes occur instantaneously and then persist. This assumption is unlikely to hold for our data, where changes may occur gradually or at any point within the given intervals, as illustrated by the solid and dashed bands. Moreover, in paired series, the timing of distributional changes may not coincide, suggesting that results should be represented as intervals rather than discrete points. Third, overlapping intervals where change points are likely to occur indicate potential associations between passive sensing variables and stress levels. However, given the observed patterns, such relationships cannot be adequately captured by models assuming a fixed functional form. Finally, as shown in Figure \ref{fig:patterns}, the low strength of signals in both series further complicates the detection of change points.

\subsection{Contributions}

In this article, we develop an algorithm that tracks simultaneous and potentially asynchronous changes in low-level statistical features, focusing on changes in the mean and variance of passive sensing data and stress levels. We define a hotspot as a time interval in which distributional changes in stress levels and passive sensing variables occur simultaneously or in temporal proximity. These features are easy to interpret and play an important role in prediction. The algorithm serves as a tool for clinical experts to investigate associations between stress and human behavior by monitoring changes in these time series.

This study makes several novel contributions. First, it introduces the concept of hotspots and proposes a data-driven method to identify them. This addresses a key limitation of conventional CPD algorithms, which typically detect only isolated time points rather than extended periods of change. By tracking local variations in statistical features, the algorithm can detect more candidate time points corresponding to actual, though possibly subtle, shifts in stress levels. Second, it jointly tracks changes across multiple distributional features, capturing diverse types of concurrent changes in a given series. This ensemble approach provides a more complete characterization of underlying dynamics than univariate metrics alone. Finally, the method is model-free: it does not assume a specific functional relationship between passive sensing variables and stress levels, while still accounting for their interdependent dynamics. The dependence structure is included as an auxiliary component, supporting detection of both within-series and cross-series changes in means and variances.

\subsection{Organization of Paper}

The remainder of this article is organized as follows. Section \ref{sec:cpd} briefly describes the definition of hotspots. Section \ref{sec:cpd} focuses on detecting change points for various types of changes. This involves the modification of the existing CPD algorithm applied in our proposed method. Section \ref{sec:hotspots_identification} describes how the initially identified change points are aggregated using ensemble strategies to generate robust change regions. The numerical experiments evaluating the performance of the method are presented in Section \ref{sec:numerical}. Finally, Section \ref{sec:discuss} concludes with a discussion.

%%%%%%%%%%%%%%%%%%%%%%%%%%%%%%%%%%%%%%%%%%%%%%%%%%%
\section{Definition of Hotspots}\label{sec:hotspots_definition}

Hotspots are defined as collections of time intervals during which clinical experts are advised to pay attention to changes in features derived from passive sensing variables. These features capture aspects of human behavior (e.g., mobilization, sociability, and sleep), and changes in these behaviors around specific time points may be strongly associated with variations in stress levels. The intervals are characterized by distributional shifts in both passive sensing data and stress levels, under the assumption that variations in one series may influence or be influenced by variations in the other.

The proposed algorithm proceeds in two main steps. First, it identifies specific time points at which changes are likely to occur, capturing different types of changes in the series. Second, for each type of detected change, it constructs neighborhood intervals around these points and combines them to form an ensemble of change regions. Consequently, the proposed method offers a flexible framework for identifying periods of behavioral and stress-related changes that are likely to be clinically meaningful.

%%%%%%%%%%%%%%%%%%%%%%%%%%%%%%%%%%%%%%%%%%%%%%%%%%%
\section{Finding Change Points}\label{sec:cpd}

As the first step of identifying hotspots, we find the change points where all different types of distributional shifts occur in either univariate or bivariate series. 

Throughout the rest of the article, we denote a univariate passive sensing variable by $X_t$ and stress levels by $Y_t$, where $\{X_t\}$ and $\{Y_t\}$ refer to subsets or complete sets of samples of $X_t$ and $Y_t$, respectively.

\subsection{Pre-processing}\label{ssec:preprocessing}

To apply the proposed algorithm to the raw dataset, two pre-processing steps are necessary. First, we handle the discrete nature of the stress-level series by transforming them into continuous representations, enabling the use of statistical models and change point detection methods that assume continuous inputs. Second, we debias the passive sensing variables by distinguishing genuinely low behavioral activity from spuriously low measurements caused by device non-use (e.g., when the phone is turned off, not carried, or the app is inactive) and applying a bias-correction procedure based on auxiliary device-use information.

In many cases, the stress-level series $\{Y_t\}$ is not continuous but instead takes discrete values following a categorical distribution, such as a 5-point Likert scale (see second or fourth row in Figure \ref{fig:patterns}). To facilitate statistical modeling and analysis that assume continuity, it is often useful to transform such discrete-valued data into a continuous form. Inspired by the treatment of quantile residuals \cite[e.g.,][]{dunn1996randomized}, we apply a smoothed and randomized inverse cumulative distribution function (CDF) transformation. Specifically, for each observation $Y_t=y$, $y\in\{1,\ldots,5\}$, we define
\begin{equation}\label{e:transformation} 
    \begin{array}{ll} U_t &:= \hat{F}_{t}(Y_t^{-}) + W_t\hat{\mathbb{P}}(Y_t=y), \\ Z_t &:= \Phi^{-1}(U_t) 
    \end{array} 
\end{equation} 
where $\hat{F}_{t}(\cdot)$ denotes the estimated CDF of the observed discrete-valued series $\{Y_t\}_{t=1}^{n}$, with $\hat{F}_{t}(y^{-})=\hat{\mathbb{P}}(Y_t<y)$ denoting the cumulative probability immediately below category $y$. $W_t\sim\mathrm{Unif}(0,1)$ are independent random variables drawn at each time point, and $\Phi^{-1}(\cdot)$ represents the inverse CDF of the standard normal distribution. This transformation effectively maps the discrete categorical responses into continuous latent variables while preserving their probabilistic structure (see the top panels of Figures \ref{fig:hotspot_mean} and Figures \ref{fig:hotspot_var} below for the examples of the transformation \eqref{e:transformation}). In what follows, we regard $\{Y_t\}$ as a continuous-valued data stream throughout the rest of this paper.

Raw passive sensing mHealth data often exhibit a downward bias due to device non-use. The 2SpamH algorithm \citep{zhang20242spamh} is a two-stage unsupervised pre-processing method developed to mitigate this bias in passively collected mHealth data. In the first stage, the algorithm evaluates the reliability of each observation using auxiliary indicators such as screen unlock counts, battery variance, and app usage, allowing it to distinguish periods of true low activity from those resulting from device non-use. In the second stage, it applies bias-correction models to the subset of reliable data, ensuring that the retained measurements more accurately capture the user’s actual behavior. Further details of the algorithm are provided in \citet{zhang20242spamh}. This pre-processing has already been applied to the real data used in Section \ref{sec:data_appl}.

\subsection{Finding Simultaneous Change Points in Univariate Series}\label{ssec:cpd_algorithm_univ}

The moving sum \cite[MOSUM; e.g.,][]{bauer1980extension, chu1995mosum,huvskova1990asymptotics} is a widely used CPD procedure based on scan statistics \cite[e.g.,][]{abolhassani2021up}; by sliding the window across the series, the procedure quantifies the difference between the statistics computed from the observations before and after the window center to evaluate shifts in mean levels. Despite its simplicity, it has received less attention compared to global search methods using dynamic programming \cite[e.g.,][]{jackson2005algorithm,killick2012optimal} or other rolling-window-based estimators, such as binary segmentation \cite[e.g.,][]{vostrikova1981detecting,fryzlewicz2014wild}. With further developments in scan-statistics-based CPD algorithms \cite[e.g.,][]{chen2015graph} and the establishment of statistical properties of the MOSUM procedure \cite{eichinger2018mosum}, it has regained attention. More recently, extensions of the MOSUM approach have been proposed to detect changes in variance \citep{messer2018multiple}, as well as the joint detection of mean and variance changes, referred to as Joint-MOSUM \citep{meier2021mosum,levajkovic2023multiscale}, to detect changes in both the mean and variance of individual series.

Specifically, suppose the series $\{X_t\}_{t=1}^n$ is given. The Joint-MOSUM algorithm identifies the time points at which changes are likely to occur. Let $\mu_X(t_1, t_2)$ and $\sigma^2_X(t_1,t_2)$ denote the population mean and variance of the observations over the interval $t_1$ and $t_2$, and their sample analogues
\begin{displaymath}
    \bar{X}(t_1,t_2)=\frac{1}{t_2-t_1+1}\sum_{t=t_1}^{t_2} X_t, \quad 
    S_X^2(t_1,t_2)= \frac{1}{t_2-t_1+1}\sum_{t=t_1}^{t_2}(X_t - \bar{X}(t_1,t_2))^2.
\end{displaymath}
In this setting, the CPD algorithm is equivalent to performing a combined hypothesis test sequentially; for each time $t=k$, we test the following null hypothesis against its alternative:
\begin{equation}\label{e:hypotheses_joint}
    H_0(k): \mu_X(k-G+1,k) = \mu_X(k+1,k+G), \ \sigma_X^2(k-G+1,k) = \sigma_X^2(k+1,k+G), 
\end{equation}
with the corresponding alternative hypotheses $H_A(k)$ defined as the negation of $H_0(k)$, indicating that either or both of the features have changed locally at time $t=k$. Here, $G$ denotes the half-window size, so that the local features are computed using $G$ observations on each side of $k$. The choice of $G$ determines the temporal scale of detection; smaller values are more sensitive to closely spaced changes, whereas larger values provide greater stability for detecting more persistent changes. In our data analysis using daily observations, we consider $G=20$ and $40$, corresponding to local windows of 20 and 40 days on each side of a candidate change point. For notational simplicity, we suppress $G$ in these features and their sample analogs hereafter. The same hypothesis test is performed for $\{Y_t\}_{t=1}^n$.

Following the conventional MOSUM procedure, we first consider changes in the local mean of $X_t$ around each candidate time point $t=k$. For an interior time point $k=G,\ldots,n-G$, there are $G$ observations available on both sides of $k$, and the local change is measured by comparing the means of these two adjacent windows. Near the beginning or end of the series, however, a complete window is not available on both sides, and boundary-adjusted statistics are used instead. Specifically,
\begin{equation}\label{e:difference_mean}
    d\bar{X}(k) 
    = \left\{\begin{array}{ll} 
     C_1\sum_{t=1}^k(\bar{X}(1,2G)-X_{t}), & k = 1,\ldots,G-1, \\
    \bar{X}(k+1,k+G) - \bar{X}(k-G+1,k), & k=G,\ldots,n-G, \\
    C_n\sum_{t=n-G+1}^n(\bar{X}(n-2G+1,n)-X_{t}), & k = n-G+1,\ldots,n,
    \end{array}\right.
\end{equation}
where $C_1=\frac{2}{\sqrt{k(2G-k)}}$ and $C_n=\frac{2}{\sqrt{(n+1-k)(k-n+2G)}}$. The first and third cases correspond to candidate change points within $G$ observations of the left and right boundaries, respectively, whereas the second case corresponds to interior candidate points for which two complete windows of length $G$ can be compared. The factors $C_1$ and $C_n$ are boundary normalization constants used in the MOSUM construction to adjust for the smaller and unequal effective sample sizes near the endpoints \citep{meier2021mosum}.

Using this local mean difference, we define the detector for a change in the mean of $X_t$ around $t=k$ as
\begin{equation}\label{e:detector_mean}
    T_{X}^{(1)}(k) = \sqrt{\frac{G}{2}}\frac{d\bar{X}(k)}{\bar{S}_{X}(k)},
\end{equation}
where $\bar{S}_{X}^2(k)$ is the corresponding local variance estimator,
\begin{equation}\label{e:bar_S}
    \bar{S}_{X}^2(k) = \left\{\begin{array}{ll}
        S_X^2(1,2G), & k = 1,\ldots,G-1, \\
        \frac{1}{2} (S_{X}^2(k+1,k+G)+S_{X}^2(k-G+1,k)), & k=G,\ldots,n-G, \\
        S_X^2(n-2G+1,n), & k = n-G+1,\ldots,n.
    \end{array}\right.
\end{equation}
Thus, for interior points, $\bar{S}_{X}^2(k)$ averages the variability in the two windows being compared, while near the boundaries it is estimated from the corresponding $2G$ observations used for boundary correction. Larger values of $|T_X^{(1)}(k)|$ therefore indicate stronger evidence of a local mean change around $k$.

The same construction is applied to the stress-level series $\{Y_t\}$. In particular, the same window size $G$ and the same boundary adjustments are used to define $d\bar{Y}(k)$ and $\bar{S}_Y(k)$, yielding
\begin{displaymath}
    T_{Y}^{(1)}(k) = \sqrt{\frac{G}{2}}\frac{d\bar{Y}(k)}{\bar{S}_{Y}(k)}.
\end{displaymath}
Using a common $G$ for $X_t$ and $Y_t$ ensures that local changes in the two series are evaluated over the same temporal scale when they are subsequently combined to identify hotspots.

The studies by \cite{messer2018multiple, messer2022bivariate} and \cite{levajkovic2023multiscale} have shown that this construction can also be applied to detect changes in variance. Specifically, the difference in the local variance of $X_t$ around the time point $t=k$, $k=1,\ldots,n$, is defined as
\begin{displaymath}
    dS_{X}^2(k) 
    = \left\{\begin{array}{ll} 
    C_1\sum_{t=1}^k(S_{X}^2(1,2G)-(X_{t}-\bar{X}(1,2G))^2)    & k = 1,\ldots,G-1, \\
    S_{X}^2(k+1,k+G) - S_{X}^2(k-G+1,k) & k=G,\ldots,n-G, \\
    C_n\sum_{t=n-G+1}^n(S_{X}^2(n-2G+1,n)-(X_{t}-\bar{X}(n-2G+1,n))^2)    & k = n-G+1,\ldots,n,
    \end{array}\right.
\end{displaymath}
using the same constants $C_1$ and $C_n$ defined in \eqref{e:difference_mean}. The rolling window estimator for the local change in the variance of $\{X_t\}$ around $t=k$ is then defined as
\begin{equation}\label{e:detector_var}
    T_{X}^{(2)}(k) = \sqrt{\frac{G}{2}}\frac{dS_{X}^2(k)}{\bar{V}_{X}(k)},
\end{equation}
where $\bar{V}_X^2(k)$ is the locally averaged estimator of the variance of the squared and centered observations at $t=k$, given by
\begin{equation}\label{e:bar_V}
    \bar{V}_{X}^2(k) = \left\{\begin{array}{ll} 
    V_X^2(1,2G), & k = 1,\ldots,G-1, \\
    \frac{1}{2}(V_{X}^2(k+1,k+G)+V_{X}^2(k-G+1,k)), & k=G,\ldots,n-G, \\
    V_X^2(n-2G+1,n), & k = n-G+1,\ldots,n,
    \end{array}\right.
\end{equation}
where $V_{X}^2(a,b) = \frac{1}{b-a+1}\sum_{t=a}^b\left((X_t - \bar{X}(a,b))^2-S_{X}^2(a,b)\right)^2$, using the locally averaged mean and variance defined in \eqref{e:bar_S}. The rolling window estimator for local changes in the variance of $\{Y_t\}$ around $t=k$, denoted by $T_Y^{(2)}(k)=\sqrt{\frac{G}{2}}dS_{Y}^2(k)/\bar{V}_{Y}(k)$ is defined similarly.

To test the hypothesis \eqref{e:hypotheses_joint}, following \cite{messer2022bivariate} and \cite{levajkovic2023multiscale}, we combine the mean- and variance-change detectors in \eqref{e:detector_mean} and \eqref{e:detector_var}. The joint detector at time point $t=k$ is defined as
\begin{displaymath}
    J_{X}(k) = \begin{pmatrix}
        T_{X}^{(1)}(k) \\ T_{X}^{(2)}(k)
    \end{pmatrix}\in\mathbb{R}^2.
\end{displaymath}
The two components of $J_X(k)$ are generally correlated because local changes in the sample mean and variance are computed from the same observations. Under the null hypothesis, implying that there is no change point, their dependence is characterized by the population correlation matrix
\begin{displaymath}
    \Gamma_{X} 
    = \begin{pmatrix}
    1 & \rho_{X} \\ \rho_{X} & 1
    \end{pmatrix},
\end{displaymath}
where $\rho_X$ is the population correlation between $X-\mu_X$ and $(X-\mu_X)^2$ which are invariant to time, given by
\begin{displaymath}
    \rho_{X} 
    := \frac{\mathbb{E}[(X-\mu_X)^3]}{\sqrt{\textrm{Var}(X-\mu_X)}\sqrt{\textrm{Var}((X-\mu_X)^2)}}
    = \frac{\kappa_{X}}{\sigma_X \nu_X}.
\end{displaymath}
Its local sample analogue around $t=k$, denoted by $\hat{\Gamma}_{X}(k)$, is obtained by replacing $\rho_X$ with its local estimate $\hat{\rho}_{X}(k)$, where
\begin{equation}\label{e:hat_corr}
    \hat{\rho}_{X}(k) = \frac{\bar{K}_{X}(k)}{\bar{S}_{X}(k)\bar{V}_{X}(k)},
\end{equation}
with 
\begin{displaymath}
    \bar{K}_{X}(k)=\frac{1}{2}(K_{X}(k+1,k+G)+K_{X}(k-G+1,k)), \quad k=G,\ldots,n-G,
\end{displaymath}
and
\begin{displaymath}
    K_{X}(a,b)=\frac{1}{b-a+1}\sum_{t=a}^b(X_t-\bar{X}(a,b))^3,
\end{displaymath}
and define boundary versions of $\bar{K}_{X}(k)$ analogous to $\bar{S}_X^2(k)$ in \eqref{e:bar_S}. Thus, $\hat{\Gamma}_{X}(k)$ accounts for the local dependence between the mean- and variance-change detectors without requiring an explicit parametric model for that dependence.

Since the two components of $J_X(k)$ may be correlated, simply combining them through their squares would ignore their dependence and could overstate changes occurring along highly correlated directions. Thus, we use the Mahalanobis distance, which rescales the joint detector by its local covariance structure and provides a measure of the magnitude of the combined departure from the null hypothesis. Specifically, 
\begin{equation}\label{e:mahalanobis_joint}
    D_{X}(k) = \sqrt{J_{X}(k)'\hat{\Gamma}_{X}(k)^{-1}J_{X}(k)}.
\end{equation}
When the two detector components are uncorrelated, $\hat{\Gamma}_{X}(k)$ is approximately the identity matrix and $D_X(k)$ reduces to the squared Euclidean norm of $J_X(k)$. In general, larger values of $D_X(k)$ indicate stronger evidence that the local mean, variance, or both have changed around $t=k$. We therefore use $D_X^2(k)$ as the test statistic at each $k=1,\ldots,n$.

The Joint-MOSUM procedure identifies initial change points where the corresponding detectors \eqref{e:mahalanobis_joint} exceed the threshold determined by $n,G$, and $\alpha$. However, no closed-form expression of the threshold exists \cite[e.g., see Section 5 in][]{messer2022bivariate}. Instead, we use the Monte Carlo simulation to compute the threshold. Specifically, for each $b^{\textrm{th}}$ replication, $b=1,\ldots,B$, the threshold is computed as
\begin{displaymath}
   D_n(\mathcal{G},\alpha)^{(b)} = \max_{g\in\mathcal{G}}\max_{g \leq h \leq n-g} \left\{ T^{(b)}(h)=\sqrt{(T_1(h)^{(b)})^2+(T_2(h)^{(b)})^2} \right\},
\end{displaymath}
where $T_m(h)^{(b)}=\frac{W_m^{(b)}(h+g)-2W_m^{(b)}(h)+W_m^{(b)}(h-g)}{\sqrt{2g}}$ for $m =1,2$, and $W_m^{(b)}(t)$ denotes a standard random walk at time $t$. The bandwidth set $\mathcal{G}$ is defined as a grid ranging from 25 up to $\min((n-1)/2,200)$. We use $B=1000$. Then the threshold $D_n(\mathcal{G},\alpha)$ is determined as the $100(1 - \alpha)^{\textrm{th}}$ percentile of $\{D_n(\mathcal{G},\alpha)^{(b)}\}$, which is compared to the Mahalanobis distance in \eqref{e:mahalanobis_joint}. The initial change points (called initial candidates) are thus identified by collecting all $k$ such that
\begin{displaymath}
   D_{X}(k) > D_n(\mathcal{G},\alpha).
\end{displaymath}

Finally, from the initial set of candidate points, we determine the finalized change points. MOSUM-type approaches are known to often detect an excessive number of change points due to the scaling (i.e., $\bar{S}_X(k)$) of the detectors \cite[e.g.,][]{eichinger2018mosum, meier2021mosum}. To mitigate over-detection, the final change points are selected as the local maxima of the following expression for a small $\eta\in(0,1)$:
\begin{displaymath}
   \hat{k} = \argmax_{\bar{k}-\eta G \leq k \leq \bar{k}+\eta G} D_{X}(k),
\end{displaymath}
where $\bar{k}$ is an initial candidate satisfying the threshold criterion. Here, $\eta G$ determines the size of the neighborhood over which nearby candidate points are consolidated into a single local maximum. A smaller $\eta$ uses a narrower neighborhood and therefore retains more distinct nearby peaks, increasing detection sensitivity by improving the ability to identify closely spaced changes, although potentially at the cost of more false or redundant detections. Conversely, a larger $\eta$ more aggressively merges nearby candidates and thus reduces over-detection (see Section 2.2 in \cite{meier2021mosum} for a related discussion). In practice, we set $\eta=0.2$ when numerous or closely spaced change points are expected, as in our simulation settings. The same procedure is applied to the other series $\{Y_t\}$.

\subsection{Finding Simultaneous Change Points in Bivariate Series}\label{ssec:cpd_algorithm_bi}

Using the Joint-MOSUM approach described in Section \ref{ssec:cpd_algorithm_univ}, we extend the method to detect simultaneous changes across multiple series, referred to as Bi-MOSUM. Applying Joint- and Bi-MOSUM to $\{X_t\}$ and $\{Y_t\}$ and combining the resulting detections allows us to identify different types of combined cross-changes. It is worth noting that Joint- and Bi-MOSUM-type algorithms can be readily extended to detect other types of changes (e.g., in autocovariances), which is beyond the scope of the study.

Suppose that a pair of series $\{Y_t\}_{t=1}^n$ and $\{X_t\}_{t=1}^n$ is given. At each candidate time point $t=k$, we consider four possible combinations of local distributional changes across the two series. Specifically, we test
\begin{equation}\label{e:hypotheses_bi}
    \begin{array}{c}
    H_0^{(\mu_Y,\mu_X)}(k): \mu_Y(k-G+1,k) = \mu_Y(k+1,k+G), \ \mu_X(k-G+1,k) = \mu_X(k+1,k+G), \\
    H_0^{(\mu_Y,\sigma_X^2)}(k): \mu_Y(k-G+1,k) = \mu_Y(k+1,k+G), \ \sigma_X^2(k-G+1,k) = \sigma_X^2(k+1,k+G), \\
    H_0^{(\sigma_Y^2,\mu_X)}(k): \sigma_Y^2(k-G+1,k) = \sigma_Y^2(k+1,k+G), \ \mu_X(k-G+1,k) = \mu_X(k+1,k+G),\\
    H_0^{(\sigma_Y^2,\sigma_X^2)}(k): \sigma_Y^2(k-G+1,k) = \sigma_Y^2(k+1,k+G), \ \sigma_X^2(k-G+1,k) = \sigma_X^2(k+1,k+G),
    \end{array}
\end{equation}
The corresponding alternative hypothesis for each case is the negation of its null hypothesis. Thus, $H_{0}^{(\mu_Y,\mu_X)}(k)$ tests whether the local means of both $Y_t$ and $X_t$ remain unchanged across the two windows adjacent to $k$, while $H_{0}^{(\mu_Y,\sigma_X^2)}(k)$ tests for stability of the local mean of $Y_t$ and the local variance of $X_t$. Similarly, $H_{0}^{(\sigma_Y^2,\mu_X)}(k)$ concerns the local variance of $Y_t$ and the local mean of $X_t$, and $H_{0}^{(\sigma_Y^2,\sigma_X^2)}(k)$ concerns the local variances of both series. Rejection of any one of these hypotheses indicates that at least one of the corresponding features differs between the $G$ observations before and after $k$, representing a local cross-change around that time point. Again, we suppress the window size $G$ in the population and sample feature representations hereafter.

Specifically, consider the pair consisting of the local mean of $Y_t$ and the local variance of $X_t$ as an example. The joint detector for this combination of changes at time $t=k$ is defined as
\begin{displaymath}
    J_{YX^2}(k) = \begin{pmatrix}
        T_{Y}^{(1)}(k) \\ T_{X}^{(2)}(k)
    \end{pmatrix}\in\mathbb{R}^2.
\end{displaymath}
The two components of $J_{YX^2}(k)$ may be correlated because they are computed from paired observations. Under the null hypothesis, their time-invariant dependence is characterized by
\begin{displaymath}
\Gamma_{YX^2} = \begin{pmatrix}
                1 & \rho_{YX^2} \\ \rho_{YX^2} & 1
                \end{pmatrix},
\end{displaymath}
where
\begin{displaymath}
    \rho_{YX^2} 
    := \frac{\mathbb{E}[(Y-\mu_Y)(X-\mu_X)^2]}{\sqrt{\textrm{Var}(Y-\mu_Y)}\sqrt{\textrm{Var}((X-\mu_X)^2)}}
    = \frac{\sigma_{YX^2}}{\sigma_Y \nu_X}.
\end{displaymath}
Its local sample analog around the time point $t=k$, denoted by $\hat{\Gamma}_{YX^2}(k)$, is computed as
\begin{equation*}
    \hat{\rho}_{YX^2}(k) = \frac{\bar{\sigma}_{XY^2}(k)}{\bar{S}_{Y}(k)\bar{V}_{X}(k)},
\end{equation*}
where $\bar{S}_{Y}(k)$ and $\bar{V}_{X}(k)$ are locally averaged variance and variance of the squared and centered observations defined in \eqref{e:bar_S} and \eqref{e:bar_V}, respectively, and $\bar{\sigma}_{YX^2}(k)=\frac{1}{2}(\hat{\sigma}_{YX^2}(k+1,k+G) + \hat{\sigma}_{YX^2}(k-G+1,k))$ denotes the locally averaged estimated cross-covariance at $t=k$, with
\begin{displaymath}
    \hat{\sigma}_{YX^2}(a,b) = \frac{1}{b-a+1}\sum_{t=a}^b(Y_t-\bar{Y}(a,b))(X_t-\bar{X}(a,b))^2.
\end{displaymath}
Thus, this construction accounts for the local dependence between the mean-related component of $Y_t$ and the variance-related component of $X_t$. Finally, the Mahalanobis distance used to detect the combined local change at time $t=k$ is computed as
\begin{equation}\label{e:mahalanobis_bi}
    D_{YX^2}(k) = \sqrt{J_{YX^2}(k)'\hat{\Gamma}_{YX^2}(k)^{-1}J_{YX^2}(k)}.
\end{equation}
Note that we can also consider the other three combinations of local cross-distributional shifts described in \eqref{e:hypotheses_bi}, where the corresponding Mahalanobis distances are denoted as $D_{YX}(k),D_{Y^2X}(k)$, and $D_{Y^2X^2}(k)$, $k=1,2,\ldots,n$, defined analogously to \eqref{e:mahalanobis_bi}.

Analogous to the Joint-MOSUM procedure, the Mahalanobis distance in \eqref{e:mahalanobis_bi} is compared with the same threshold $D_n(\mathcal{G},\alpha)$ 
\begin{displaymath}
   D_{YX^2}(k) > D_n(\mathcal{G},\alpha), \quad k=1,\ldots,n.
\end{displaymath}
to determine whether the time points $t=k$, $k=1,\ldots,n$, are initial candidate change points. Finally, the same screening rule as in the Joint-MOSUM procedure is applied; for a small $\eta\in(0,1)$: we obtain
\begin{displaymath}
   \hat{k} = \argmax_{\bar{k} -\eta G \leq k \leq \bar{k}+\eta G} D_{YX^2}(k).
\end{displaymath}
where $\bar{k}$ is an initial candidate. The remaining three combined hypothesis tests for local cross-changes are performed similarly.

%%%%%%%%%%%%%%%%%%%%%%%%%%%%%%%%%%%%%%%%%%%%%%%%%%%
\section{Identifying Hotspots}\label{sec:hotspots_identification}

We construct hotspots as intervals around the identified time points and merge them into regions that capture both simultaneous and possibly asynchronous changes across the two series. By highlighting periods in which distributional changes in the two series occur at the same or nearby times, the resulting hotspots identify potential temporal associations between them without requiring an explicit parametric model or their relationship formulation.

\subsection{Thresholding Rule}

We identify sets of time points for which the corresponding Mahalanobis distances exceed the given threshold; consecutive selected time points form hotspot intervals.

Continuing from the previous example, consider the detector that combines changes in the mean of $\{Y_t\}$ and the variance of $\{X_t\}$. Define the set of time points at which this cross-series detector exceeds the threshold as
\begin{equation}\label{e:rule_threshold_YX2}
    \hat{\mathcal{H}}_{\mathrm{Thr},YX^2} 
    := \left\{k\in\{1,\ldots,n\}: D_{YX^2}(k)>D_n(\mathcal{G},\alpha)\right\}.
\end{equation}
Similarly, for the Joint-MOSUM detector applied to $\{Y_t\}$, define
\begin{equation}\label{e:rule_threshold_Y}
    \hat{\mathcal{H}}_{\mathrm{Thr},Y} 
    := \left\{k\in\{1,\ldots,n\}: D_Y(k)>D_n(\mathcal{G},\alpha) \right\}.
\end{equation}
The hotspot under the thresholding rule is then defined by the intersection of \eqref{e:rule_threshold_YX2} and \eqref{e:rule_threshold_Y},
\begin{equation}\label{e:rule_threshold_YX2_Y}
    \hat{\mathcal{H}}_{\mathrm{Thr}} := \hat{\mathcal{H}}_{\mathrm{Thr},YX^2} \cap \hat{\mathcal{H}}_{\mathrm{Thr},Y}.
\end{equation}
Thus, a time point belongs to $\hat{\mathcal{H}}_{\mathrm{Thr}}$ only when both the selected cross-series detector and the Joint-MOSUM detector for $\{Y_t\}$ exceed the common threshold.

Other combinations of the four cross-series Mahalanobis distances,
$D_{YX}(k)$, $D_{YX^2}(k)$, $D_{Y^2X}(k)$, and $D_{Y^2X^2}(k)$, can be considered depending on the types of changes of interest. When no particular type of cross-change is specified, we define
\begin{equation}\label{e:rule_threshold_full}
    \hat{\mathcal{H}}_{\mathrm{Thr,full}} 
    := \left\{ k\in\{1,\ldots,n\}: \max\left( D_{YX}(k), D_{YX^2}(k), D_{Y^2X}(k), D_{Y^2X^2}(k) \right) > D_n(\mathcal{G},\alpha) \right\}.
\end{equation}
In this case, the corresponding hotspot is obtained by replacing $\hat{\mathcal{H}}_{\mathrm{Thr},YX^2}$ in \eqref{e:rule_threshold_YX2_Y} with $\hat{\mathcal{H}}_{\mathrm{Thr,full}}$ defined in \eqref{e:rule_threshold_full}, yielding
\begin{equation}\label{e:rule_threshold_full_Y}
    \hat{\mathcal{H}}_{\mathrm{Thr}} 
    = \hat{\mathcal{H}}_{\mathrm{Thr,full}} \cap \hat{\mathcal{H}}_{\mathrm{Thr},Y}.
\end{equation}
More generally, $\hat{\mathcal{H}}_{\mathrm{Thr},YX^2}$ in \eqref{e:rule_threshold_YX2} may be replaced by a set defined using any subset of the four cross-series detectors, depending on the distributional changes under consideration. We refer to this construction as the thresholding rule.

\subsection{Confidence Interval Rule}

CPD algorithms are designed to detect abrupt shifts rather than gradual transitions. Because our goal is to construct hotspot intervals rather than report isolated change point estimates, we augment the estimated change points with intervals that quantify uncertainty around their locations. While extensive research has focused on developing CPD algorithms, relatively few studies have addressed uncertainty quantification, aside from a few Bayesian approaches \cite[e.g.,][]{belitser2025bayesian,cappello2025bayesian}. In frequentist approaches, methods based on controlling the family-wise error rate \citep{frick2014multiscale}, false discovery rate \citep{li2016fdr}, and post-selection inference \citep{hyun2016exact} have been proposed. However, these approaches are not directly designed to produce the interval-valued representation of change point uncertainty required for our hotspot construction, as illustrated in Figure \ref{fig:patterns}.

Recently, \cite{cho2022bootstrap} proposed two methods for constructing confidence intervals for change points in univariate MOSUM: one based on pointwise coverage and the other on controlling the family-wise error rate. However, our empirical results show that the latter often produces overly narrow intervals, which is undesirable in our context where overlapping regions between different types of changes need to be preserved. Therefore, we adopt the pointwise method in our framework.

Confidence intervals for change points from the Joint- and Bi-MOSUM procedures can be constructed analogously to those for univariate MOSUM change points, following the bootstrap approach in Section 2.3 of \cite{cho2022bootstrap} and Section 2.7 of \cite{meier2021mosum}. For the Bi-MOSUM procedure, the paired observations $(Y_t,X_t)$ are resampled jointly within each estimated segment.

Continuing from the previous example, suppose there are $\hat{K}_{YX^2}=|\hat{\mathcal{T}}_{YX^2}|$ estimated change points associated with the detector combining the mean of $Y_t$ and the variance of $X_t$. We order them as
\begin{displaymath}
    0 = \hat{k}_{YX^2,0} < \hat{k}_{YX^2,1} < \ldots < \hat{k}_{YX^2,\hat{K}_{YX^2}} < \hat{k}_{YX^2,\hat{K}_{YX^2}+1} = n.
\end{displaymath}
Define the corresponding segments as
\begin{displaymath}
    I_{YX^2,j}=\{\hat{k}_{YX^2,j-1}+1,\ldots,\hat{k}_{YX^2,j}\}, \quad j=1,\ldots,\hat{K}_{YX^2}+1.
\end{displaymath}
Within each segment, we draw a random sample of size $|I_{YX^2,j}|$ with
replacement from the paired observations $\{(Y_t,X_t):t\in I_{YX^2,j}\}$. Resampling the two series jointly preserves the contemporaneous dependence between $Y_t$ and $X_t$ within each estimated segment. Repeating this procedure across all segments yields bootstrap samples
\begin{displaymath}
    \left\{(Y_{1}^{(b)},X_{1}^{(b)}),\ldots,(Y_{n}^{(b)},X_{n}^{(b)})\right\}, \quad b=1,\ldots,B.
\end{displaymath}
For each bootstrap replication, the $j^{\textrm{th}}$ change point is re-estimated locally as
\begin{displaymath}
    \hat{k}_{YX^2,j}^{(b)} 
    = \argmax_{\hat{k}_{YX^2,j}-G \leq k \leq \hat{k}_{YX^2,j}+G}D_{YX^2}^{(b)}(k).
\end{displaymath}
For each original estimate $\hat{k}_{YX^2,j}$, let $M_{YX^2,j}(\alpha)$ denote the $100(1-\alpha)$th percentile of 
\begin{displaymath}
    \left\{ |\hat{k}_{YX^2,j}^{(1)}-\hat{k}_{YX^2,j}|, \ldots, |\hat{k}_{YX^2,j}^{(B)}-\hat{k}_{YX^2,j}| \right\}.
\end{displaymath}
The resulting pointwise $100(1-\alpha)\%$ confidence interval is
\begin{displaymath}
    \textrm{CI}_{YX^2,j}(\alpha) 
    = \left[\max\left(1,\hat{k}_{YX^2,j} -M_{YX^2,j}(\alpha)\right),
    \min\left(\hat{k}_{YX^2,j} +M_{YX^2,j}(\alpha),n\right)\right],\quad j=1,\ldots,\hat{K}_{YX^2}.
\end{displaymath}

Define the union of these confidence intervals as
\begin{equation}\label{e:rule_confidence_YX2}
    \hat{\mathcal{H}}_{\mathrm{CI},YX^2}(\alpha)
    := \bigcup_{j=1}^{\hat{K}_{YX^2}} \mathrm{CI}_{YX^2,j}(\alpha).
\end{equation}
Similarly, let $\hat{K}_Y=|\hat{\mathcal{T}}_Y|$ and define
\begin{equation}\label{e:rule_confidence_Y}
    \hat{\mathcal{H}}_{\mathrm{CI},Y}(\alpha)
    := \bigcup_{j=1}^{\hat{K}_{Y}} \mathrm{CI}_{Y,j}(\alpha)
\end{equation}
denote the union of confidence intervals corresponding to the Joint-MOSUM change points detected in $\{Y_t\}$. The hotspot under the confidence interval rule is then defined as an intersection of \eqref{e:rule_confidence_YX2} and \eqref{e:rule_confidence_Y}:
\begin{equation}\label{e:rule_confidence_YX2_Y}
    \hat{\mathcal{H}}_{\mathrm{CI}}(\alpha)
    := \hat{\mathcal{H}}_{\mathrm{CI},YX^2}(\alpha) \cap \hat{\mathcal{H}}_{\mathrm{CI},Y}(\alpha).
\end{equation}
Thus, the CI-based hotspot consists of time points lying within confidence intervals associated with both the selected cross-series changes and changes detected in $\{Y_t\}$.

If all four types of cross-series changes are considered, define
\begin{align}
    & \hat{\mathcal{H}}_{\mathrm{CI,full}}(\alpha) \nonumber\\
    :=&  \left( \bigcup_{j_1=1}^{\hat{K}_{YX}} \mathrm{CI}_{YX,j_1}(\alpha)\right) \cup
    \left( \bigcup_{j_2=1}^{\hat{K}_{YX^2}} \mathrm{CI}_{YX^2,j_2}(\alpha)\right) \cup \left( \bigcup_{j_3=1}^{\hat{K}_{Y^2X}} \mathrm{CI}_{Y^2X,j_3}(\alpha)\right) \cup
    \left( \bigcup_{j_4=1}^{\hat{K}_{Y^2X^2}} \mathrm{CI}_{Y^2X^2,j_4}(\alpha)\right).  \label{e:rule_confidence_full}
\end{align}
The corresponding hotspot is obtained by replacing
$\hat{\mathcal{H}}_{\mathrm{CI},YX^2}(\alpha)$ in \eqref{e:rule_confidence_YX2_Y} with $\hat{\mathcal{H}}_{\mathrm{CI,full}}(\alpha)$ in \eqref{e:rule_confidence_full}:
\begin{equation}\label{e:rule_confidence_full_Y}
    \hat{\mathcal{H}}_{\mathrm{CI}}(\alpha)
    = \hat{\mathcal{H}}_{\mathrm{CI,full}}(\alpha) \cap \hat{\mathcal{H}}_{\mathrm{CI},Y}(\alpha).
\end{equation}
More generally, $\hat{\mathcal{H}}_{\mathrm{CI},YX^2}(\alpha)$ in
\eqref{e:rule_confidence_YX2_Y} may be replaced by the union of confidence
intervals associated with any selected subset of the four cross-series
detector combinations, depending on the distributional changes under
consideration. We refer to this construction as the confidence interval rule.

\subsection{Illustration of Algorithm}\label{ssec:illustration}

Two illustrative examples of hotspot identification are shown in Figures \ref{fig:hotspot_mean} and \ref{fig:hotspot_var}, provided in Appendix \ref{ap:appendix}.

The first example in Figure \ref{fig:hotspot_mean} shows a scenario in which the stress level series $\{Y_t\}$ undergoes a single change in its mean, while the corresponding passive sensing variable exhibits simultaneous changes in both its mean and variance, with a five–time-point lag between the two series. The second example in Figure \ref{fig:hotspot_var} demonstrates a case where the variance of the stress level changes twice, accompanied by two corresponding changes in the passive sensing variable, each occurring with a similar five–time-point lag. Furthermore, transforming the categorical stress-level series $\{Y_t\}$ into a continuous form using \eqref{e:transformation} shows that the characterization of mean and variance shifts in the categorical distributions is preserved after the transformation described in Section \ref{ssec:preprocessing}. This is illustrated in Figure \ref{fig:hotspot_mean}, which captures mean shifts, and Figure \ref{fig:hotspot_var}, which detects variance changes.

%%%%%%%%%%%%%%%%%%%%%%%%%%%%%%%%%%%%%%%%%%%%%%%%%%%
\section{Numerical Studies}\label{sec:numerical}

The numerical studies consist of three parts. First, through simulation studies, we examine whether jointly detecting multiple types of distributional changes improves change point detection when different types of changes coexist. We also compare the proposed Joint- and Bi-MOSUM approaches with benchmark CPD methods. Unlike conventional evaluations that emphasize exact localization, our primary interest is whether detected change points occur sufficiently close to the true change locations to serve as reliable anchors for subsequent hotspot construction

Second, we investigate the properties of the proposed hotspot definitions, namely the thresholding and confidence-interval rules. We evaluate whether they produce relatively concise intervals while retaining the true change points, particularly when changes in the two series occur at slightly different times. This asynchronous setting is important because requiring changes to occur at exactly the same time may fail to identify temporally related changes across stress levels and passive sensing variables; a useful hotspot should accommodate such discrepancies without becoming unnecessarily wide. Rather than ranking one rule as universally preferable, this experiment characterizes the trade-off between coverage of relevant changes and hotspot width under different data-generating settings.

Finally, we apply the proposed rules to real-world data to illustrate their practical utility and interpretability in applied settings.

\subsection{Simulation on Change Points}\label{ssec:cpd_simulation}

First, we focus on a univariate time series of interest. Throughout the simulation studies, the sample size is fixed at $n=100$, as the lengths of passive sensing variables and stress-level series are typically not sufficiently long. We consider scenarios in which the number of change points ranges from one to three, located at $k^*=50$ for a single change, $k^*=40,60$ for two changes, and $k^*=25,50,75$ for three changes. For each scenario, we examine six cases (Cases 1--6), each corresponding to a different signal strength. Case 1 represents the strongest signal, characterized by larger changes in the segment means and lower variability, whereas Case 6 represents the weakest signal. More specifically, signal strength is controlled by the magnitude of the mean shift between adjacent segments, represented by $(-\mu,\mu)$, together with the range of segment-specific standard deviations from $\sigma_{\min}$ to $\sigma_{\max}$. The parameter settings are as follows:
\begin{align*}
    \text{Case 1:} & \quad \mu=2, \sigma_{\min}=0.1, \sigma_{\max}=0.4,\\
    \text{Case 2:} & \quad \mu=2, \sigma_{\min}=0.1, \sigma_{\max}=0.8,\\
    \text{Case 3:} & \quad \mu=2, \sigma_{\min}=0.4, \sigma_{\max}=0.8,\\
    \text{Case 4:} & \quad \mu=1, \sigma_{\min}=0.1, \sigma_{\max}=0.4,\\
    \text{Case 5:} & \quad \mu=1, \sigma_{\min}=0.1, \sigma_{\max}=0.8,\\
    \text{Case 6:} & \quad \mu=1, \sigma_{\min}=0.4, \sigma_{\max}=0.8.
\end{align*}
Each setting is replicated 500 times.

We use the Joint-MOSUM method described in Section \ref{ssec:cpd_algorithm_univ} with window sizes $G=20$ and $G=40$, and compare it with several established univariate CPD methods representing different detection strategies. PELT \citep{killick2012optimal}, implemented in the R package \texttt{changepoint} \citep{killick2014changepoint}, uses a global optimization criterion with pruning to estimate multiple change points efficiently, particularly for short observation periods, whereas wild binary segmentation (WBS) \citep{fryzlewicz2014wild}, implemented in the R package \texttt{breakfast}, searches over randomly selected subintervals and is designed to detect multiple, potentially closely spaced changes. We also include \texttt{stepR} \citep{pein2017heterogeneous} as a segmentation-based method and the univariate MOSUM procedure of \citet{eichinger2018mosum}, implemented in the R package \texttt{mosum}, as a direct rolling-window comparator using the same window sizes ($G=20,40$). Together, these methods provide benchmarks spanning global optimization, recursive segmentation, and conventional MOSUM-based detection, allowing us to assess whether jointly tracking mean and variance changes provides an advantage when heterogeneous types of distributional changes coexist.

Our primary evaluation criterion is not exact localization of every change point, but whether the detected points fall sufficiently close to the true change locations to provide reliable anchors for subsequent hotspot construction. Exact timing is less critical for our purpose because hotspots are defined as intervals intended to accommodate small temporal discrepancies and potentially asynchronous changes across series. Accordingly, the simulation emphasizes the ability to recover the neighborhood of a true change while avoiding excessive false detections, rather than requiring exact recovery of the change time.

For performance evaluation, we use power and false discovery rate (FDR), defined as follows. For each replication $i=1,\ldots,500$, and let $\hat{K}^{(i)}$ denote the number of detected change points at the $i^{\textrm{th}}$iteration. Then,
\begin{align}
    \textrm{Power}(\eta) &= \frac{1}{500}\sum_{i=1}^{500}1_{\{\bigcap_{\ell=1}^{K^*}\left[\cup_{j=1}^{\hat{K}^{(i)}}\{\hat{t}_j^{(i)}-t_{\ell}^*\leq\eta\}\right]\}}, \label{e:power}\\ \textrm{FDR}(\eta) &= \frac{\sum_{i=1}^{500}1_{\{ \left\{\hat{K}^{(i)}\geq1\right\}\ \& \ \left\{\bigcap_{\ell=1}^{K^*}\left[\cap_{j=1}^{\hat{K}^{(i)}}\{|\hat{t}_j^{(i)}-t_{\ell}^*|>\eta\}\right]\right\}}}{\sum_{i=1}^{500}1_{\{\hat{K}^{(i)}\geq1\}}}, \label{e:fdr}
\end{align}
where $K^*$ denotes the number of true change points. We set $\eta=5$, so that a true change point is regarded as successfully detected if at least one estimated change point lies within five time points of its true location. Multiple detections within the same tolerance neighborhood are treated as identifying the same underlying change for this evaluation. Throughout all simulations, the significance level is fixed at $\alpha=0.05$.

Table \ref{tab:numer1_univ} presents the simulation results. When changes in the mean and variance coexist within the same time series, the Joint-MOSUM method generally achieves higher power and better FDR control than the benchmark methods. This result supports the use of a joint detector in settings where a structural change may involve the mean, the variance, or both, because methods targeting only a single distributional feature may fail to capture the full change structure. Across methods, power generally decreases, and FDR tends to increase as the signal becomes weaker, particularly when the magnitude of the mean shift decreases or the segment-specific standard deviations increase. In a few settings, performance improves slightly from Case 3 to Case 4, although this pattern is not systematic.

When there is a single change point, the larger window size ($G=40$) generally yields higher power and lower FDR. In contrast, when multiple change points are present, the shorter window ($G=20$) performs better because it is less likely to span two nearby changes within the same local window. This improves the ability to distinguish closely spaced change points, although the shorter window may also increase the number of false positive detections. In the present application, this trade-off may be acceptable because missing a potentially meaningful stress-related change is of greater concern than retaining an additional candidate change point that can subsequently be filtered during hotspot construction.

%%%%%%%%%%%%%%%%%%%%%%%%%%%%%%%%%%%%%%%%%%%%%%%%%%%
\begin{table}[t]
\resizebox{\columnwidth}{!}{%
\begin{tabular}{ccccccccccccc}
 & \multicolumn{6}{c|}{Power(5)} & \multicolumn{6}{c}{FDR(5)} \\ \hline\hline
\# of change points: 1 & Case 1 & Case 2 & Case 3 & Case 4 & Case 5 & Case 6 & Case 1 & Case 2 & Case 3 & Case 4 & Case 5 & Case 6 \\ \hline
Joint-MOSUM $(G=20)$ & \textbf{0.904} & \textbf{0.888} & \textbf{0.780} & 0.834 & 0.766 & \textbf{0.632} & 0.020 & \textbf{0.024} & 0.064 & 0.048 & 0.064 & 0.098 \\ \cline{1-1}
Joint-MOSUM $(G=40)$ & 0.910 & \textbf{0.888} & 0.762 & \textbf{0.860} & \textbf{0.810} & 0.612 & 0.022 & 0.042 & 0.088 & 0.042 & 0.090 & 0.118 \\ \cline{1-1}
PELT & 0.898 & 0.880 & 0.712 & 0.812 & 0.766 & 0.510 & 0.096 & 0.118 & 0.286 & 0.180 & 0.226 & 0.490 \\ \cline{1-1}
stepR & 0.872 & 0.822 & 0.734 & 0.768 & 0.662 & 0.528 & 0.026 & 0.058 & \textbf{0.034} & 0.050 & 0.128 & 0.064 \\ \cline{1-1}
WBS & 0.892 & 0.846 & 0.770 & 0.808 & 0.694 & 0.594 & 0.108 & 0.154 & 0.230 & 0.192 & 0.306 & 0.406 \\ \cline{1-1}
MOSUM $(G=20)$ & 0.866 & 0.798 & 0.718 & 0.748 & 0.628 & 0.508 & \textbf{0.018} & 0.026 & 0.038 & \textbf{0.036} & \textbf{0.028} & \textbf{0.052} \\ \cline{1-1}
MOSUM $(G=40)$ & 0.878 & 0.828 & 0.762 & 0.804 & 0.676 & 0.584 & 0.024 & 0.034 & 0.052 & 0.052 & 0.072 & 0.092 \\ \hline\hline
\# of change points: 2 & Case 1 & Case 2 & Case 3 & Case 4 & Case 5 & Case 6 & Case 1 & Case 2 & Case 3 & Case 4 & Case 5 & Case 6 \\ \hline
Joint-MOSUM $(G=20)$ & \textbf{0.850} & \textbf{0.756} & \textbf{0.648} & \textbf{0.674} & \textbf{0.580} & \textbf{0.380} & \textbf{0.002} & \textbf{0.008} & \textbf{0.016} & \textbf{0.012} & \textbf{0.025} & 0.066 \\ \cline{1-1}
Joint-MOSUM $(G=40)$ & 0.260 & 0.234 & 0.226 & 0.238 & 0.220 & 0.142 & 0.032 & 0.043 & 0.068 & 0.045 & 0.067 & 0.118 \\ \cline{1-1}
PELT & 0.000 & 0.000 & 0.000 & 0.000 & 0.000 & 0.000 & 0.028 & 0.070 & 0.172 & 0.096 & 0.196 & 0.452 \\ \cline{1-1}
stepR & 0.756 & 0.566 & 0.436 & 0.504 & 0.304 & 0.144 & 0.010 & 0.029 & 0.032 & 0.025 & 0.090 & 0.088 \\ \cline{1-1}
WBS & 0.824 & 0.654 & 0.594 & 0.608 & 0.442 & 0.296 & 0.016 & 0.034 & 0.062 & 0.040 & 0.140 & 0.230 \\ \cline{1-1}
MOSUM $(G=20)$ & 0.768 & 0.580 & 0.522 & 0.538 & 0.354 & 0.264 & \textbf{0.002} & 0.010 & 0.021 & 0.015 & 0.045 & \textbf{0.049} \\ \cline{1-1}
MOSUM $(G=40)$ & 0.108 & 0.088 & 0.088 & 0.100 & 0.056 & 0.056 & 0.012 & 0.025 & 0.032 & 0.031 & 0.084 & 0.106 \\ \hline\hline
\# of change points: 3 & Case 1 & Case 2 & Case 3 & Case 4 & Case 5 & Case 6 & Case 1 & Case 2 & Case 3 & Case 4 & Case 5 & Case 6 \\ \hline
Joint-MOSUM $(G=20)$ & \textbf{0.796} & \textbf{0.654} & \textbf{0.504} & \textbf{0.616} & \textbf{0.414} & \textbf{0.248} & \textbf{0.000} & \textbf{0.002} & \textbf{0.002} & \textbf{0.002} & \textbf{0.012} & \textbf{0.014} \\ \cline{1-1}
Joint-MOSUM $(G=40)$ & 0.142 & 0.136 & 0.094 & 0.126 & 0.074 & 0.034 & 0.018 & 0.040 & 0.039 & 0.034 & 0.075 & 0.109 \\ \cline{1-1}
PELT & 0.000 & 0.000 & 0.000 & 0.000 & 0.000 & 0.000 & 0.010 & 0.038 & 0.118 & 0.064 & 0.236 & 0.443 \\ \cline{1-1}
stepR & 0.602 & 0.448 & 0.236 & 0.366 & 0.162 & 0.032 & 0.004 & 0.014 & 0.022 & 0.020 & 0.045 & 0.082 \\ \cline{1-1}
WBS & 0.700 & 0.546 & 0.420 & 0.504 & 0.258 & 0.166 & 0.002 & 0.010 & 0.028 & 0.012 & 0.076 & 0.096 \\ \cline{1-1}
MOSUM $(G=20)$ & 0.682 & 0.494 & 0.386 & 0.462 & 0.212 & 0.140 & 0.002 & 0.006 & 0.006 & 0.004 & 0.026 & 0.022 \\ \cline{1-1}
MOSUM $(G=40)$ & 0.010 & 0.004 & 0.004 & 0.008 & 0.004 & 0.000 & 0.014 & 0.034 & 0.043 & 0.034 & 0.095 & 0.115 \\ \hline
\end{tabular}%
}
\caption{Power and FDR for $\eta=5$ based on simulation results using Joint-MOSUM and other benchmarks under three scenarios of a univariate series, each with six different signal strengths (Cases). A significance level of $\alpha=0.05$ is used if required for some CPD methods. The best performance in each scenario is shown in bold.}
\label{tab:numer1_univ}
\end{table}
%%%%%%%%%%%%%%%%%%%%%%%%%%%%%%%%%%%%%%%%%%%%%%%%%%%

For the next simulation studies, we ensemble the four combinations of the Bi-MOSUM results and compare them with other CPD algorithms for bivariate series. Similar to the univariate case, the sample size for each of the two series is fixed at $n=100$. We consider six different cases consistent with the univariate simulation setups. For each pair of series, distributional changes are introduced under the following scenarios: (i) a change at $k^*=40$ in the first series and $k^*=60$ in the second series, (ii) a change at $k^*=50$ in the first series and $k^*=40,60$ in the second series, and (iii) changes at $k^*=40,60$ in the first series and $k^*=30,70$ in the second series. Each setup is replicated 500 times.

For Bi-MOSUM, we run four types of combinations (mean–mean, mean–variance, variance–mean, and variance–variance) to simulate scenarios in which the types of distributional changes are not known in advance, and then merge the estimated change points. We use window sizes $G=20$ and $G=40$ for the algorithm. Although most CPD algorithms are designed for univariate data, several multivariate CPD methods have been widely validated in different contexts. Specifically, we consider the nonparametric multivariate CPD by \cite{matteson2014nonparametric}, implemented in the R package \texttt{ecp} \citep{james2015ecp}; sparsified binary segmentation by \cite{cho2015multiple}; and the double cumulative sum (CUSUM) statistic by \cite{cho2016change}, both implemented in the R package \texttt{hdbinseg}. Additionally, we include a nonparametric CPD algorithm by \cite{padilla2021optimal}, which is implemented independently of the authors$^\ddagger$\def\thefootnote{$\ddagger$}\footnotetext{The functioning code available at https://github.com/HaotianXu/changepoints/tree/main}.

Similar to the univariate simulation setup, we evaluate performance using power \eqref{e:power} and FDR \eqref{e:fdr}, with $\eta = 5$ for power and $\eta = 0$ for FDR, since allowing a tolerance for FDR would make the performance of different methods indistinguishable. Here, the true and estimated change points from both series are merged separately. For example, $K^*=3$ when there is one true change point in one series and two in the other.

Table \ref{tab:numer1_multi} presents the simulation results. The results generally mirror the patterns observed in the univariate series: as the signal strength decreases, the performance of all methods deteriorates, and more change points lead to further challenges. Nevertheless, the proposed method remains robust across all scenarios. Similar trends regarding the choice of window sizes, as seen in the univariate simulations, are also observed here. Throughout the simulation studies, the results demonstrate the necessity and effectiveness of ensembling detection results from MOSUM-type algorithms to accurately anchor the change points initially.

%%%%%%%%%%%%%%%%%%%%%%%%%%%%%%%%%%%%%%%%%%%%%%%%%%%
\begin{table}[t]
\resizebox{\columnwidth}{!}{%
\begin{tabular}{ccccccccccccc}
 & \multicolumn{6}{c|}{Power(5)} & \multicolumn{6}{c}{FDR(0)} \\ \hline\hline
\# of change points: (1,1) & Case 1 & Case 2 & Case 3 & Case 4 & Case 5 & Case 6 & Case 1 & Case 2 & Case 3 & Case 4 & Case 5 & Case 6 \\ \hline
Bi-MOSUM $(G=20)$ & 0.844 & 0.782 & 0.608 & 0.734 & 0.602 & 0.398 & 0.050 & 0.118 & 0.209 & 0.132 & 0.252 & 0.447 \\ \cline{1-1}
Bi-MOSUM $(G=40)$ & \textbf{0.968} & \textbf{0.912} & \textbf{0.830} & \textbf{0.910} & \textbf{0.772} & \textbf{0.552} & \textbf{0.024} & \textbf{0.036} & \textbf{0.090} & \textbf{0.066} & \textbf{0.101} & \textbf{0.241} \\ \cline{1-1}
ecp & 0.166 & 0.166 & 0.072 & 0.130 & 0.116 & 0.024 & 0.129 & 0.288 & 0.337 & 0.267 & 0.428 & 0.573 \\ \cline{1-1}
SBS & 0.786 & 0.688 & 0.482 & 0.606 & 0.422 & 0.208 & 0.063 & 0.110 & 0.221 & 0.152 & 0.335 & 0.458 \\ \cline{1-1}
DCBS & 0.018 & 0.028 & 0.008 & 0.018 & 0.042 & 0.022 & 0.980 & 0.935 & 0.947 & 0.953 & 0.911 & 0.880 \\ \cline{1-1}
NMP & 0.734 & 0.682 & 0.612 & 0.578 & 0.448 & 0.426 & 0.092 & 0.154 & 0.251 & 0.198 & 0.409 & 0.535 \\ \hline\hline
\# of change points: (1,2) & Case 1 & Case 2 & Case 3 & Case 4 & Case 5 & Case 6 & Case 1 & Case 2 & Case 3 & Case 4 & Case 5 & Case 6 \\ \hline
Bi-MOSUM $(G=20)$ & \textbf{0.956} & \textbf{0.842} & \textbf{0.722} & \textbf{0.818} & \textbf{0.686} & \textbf{0.436} & \textbf{0.012} & \textbf{0.032} & 0.085 & 0.048 & 0.125 & 0.289 \\ \cline{1-1}
Bi-MOSUM $(G=40)$ & 0.714 & 0.558 & 0.556 & 0.618 & 0.400 & 0.260 & \textbf{0.012} & 0.018 & \textbf{0.046} & \textbf{0.034} & \textbf{0.091} & \textbf{0.209} \\ \cline{1-1}
ecp & 0.000 & 0.002 & 0.000 & 0.000 & 0.004 & 0.000 & 0.133 & 0.236 & 0.375 & 0.315 & 0.424 & 0.544 \\ \cline{1-1}
SBS & 0.344 & 0.322 & 0.136 & 0.236 & 0.166 & 0.070 & 0.048 & 0.105 & 0.229 & 0.179 & 0.266 & 0.474\\ \cline{1-1}
DCBS & 0.000 & 0.000 & 0.000 & 0.002 & 0.000 & 0.002 & 0.964 & 0.954 & 0.915 & 0.938 & 0.902 & 0.849 \\ \cline{1-1}
NMP & 0.232 & 0.182 & 0.160 & 0.146 & 0.136 & 0.090 & 0.050 & 0.122 & 0.180 & 0.176 & 0.351 & 0.452 \\ \hline\hline
\# of change points: (2,2) & Case 1 & Case 2 & Case 3 & Case 4 & Case 5 & Case 6 & Case 1 & Case 2 & Case 3 & Case 4 & Case 5 & Case 6 \\ \hline
Bi-MOSUM $(G=20)$ & \textbf{0.928} & \textbf{0.840} & \textbf{0.696} & \textbf{0.778} & \textbf{0.590} & \textbf{0.338} & \textbf{0.002} & 0.012 & 0.028 & 0.020 & 0.068 & 0.196\\ \cline{1-1}
Bi-MOSUM $(G=40)$ & 0.431 & 0.358 & 0.282 & 0.320 & 0.204 & 0.084 & \textbf{0.002} & \textbf{0.004} & \textbf{0.026} & \textbf{0.012} & \textbf{0.046} & \textbf{0.140} \\ \cline{1-1}
ecp & 0.000 & 0.000 & 0.000 & 0.000 & 0.000 & 0.000 & 0.081 & 0.185 & 0.247 & 0.190 & 0.390 & 0.461 \\ \cline{1-1}
SBS & 0.184 & 0.078 & 0.044 & 0.068 & 0.028 & 0.004 & 0.020 & 0.089 & 0.115 & 0.092 & 0.248 & 0.391 \\ \cline{1-1}
DCBS & 0.012 & 0.022 & 0.012 & 0.010 & 0.020 & 0.016 & 0.968 & 0.928 & 0.912 & 0.932 & 0.878 & 0.872 \\ \cline{1-1}
NMP & 0.002 & 0.004 & 0.008 & 0.008 & 0.004 & 0.008 & 0.020 & 0.072 & 0.120 & 0.100 & 0.269 & 0.355 \\ \hline
\end{tabular}%
}
\caption{Power for $\eta=5$ and FDR for $\eta=0$ based on simulation results using Bi-MOSUM and other benchmarks under three scenarios of bivariate series, each with six different signal strengths (Cases). A significance level of $\alpha=0.05$ is used if required for some CPD methods. The best performance in each scenario is shown in bold.}
\label{tab:numer1_multi}
\end{table}
%%%%%%%%%%%%%%%%%%%%%%%%%%%%%%%%%%%%%%%%%%%%%%%%%%%

\subsection{Simulation on Hotspots}\label{ssec:hotspots_simulation}

We evaluate the hotspot definitions based on their ability to convey information efficiently, using shorter intervals while still capturing the true change points, particularly in the presence of asynchronous changes across series.

We consider two hotspot constructions: the thresholding rule (Thrs) and the confidence-interval rule (CI). Because the simulation does not assume prior knowledge about which distributional features change, we consider all four types of cross-series changes. Accordingly, the thresholding hotspot is defined by \eqref{e:rule_threshold_full_Y}, using $\hat{\mathcal{H}}_{\mathrm{Thr,full}}$ in \eqref{e:rule_threshold_full}, and the CI-based hotspot is defined by \eqref{e:rule_confidence_full_Y}, using $\hat{\mathcal{H}}_{\mathrm{CI,full}}(\alpha)$ in \eqref{e:rule_confidence_full}. We set $\alpha=0.05$ for both rules.

The sample size is fixed at $n=100$, and we consider the same six signal-strength cases described in Section \ref{ssec:cpd_simulation}. We examine a pair of time series, $\{X_t\}$ and $\{Y_t\}$, under scenarios containing either one or two change points. For $\{X_t\}$, the change point is located at $k^*=50$ in the one change point scenario, whereas the two change points are located at $k^*=40$ and $60$. For $\{Y_t\}$, the corresponding change points are shifted by $\delta$, occurring at $k^*=50+\delta$ and at $k^*=40+\delta$ and $60+\delta$, respectively, where $\delta\in\{0,5\}$. Thus, $\delta=0$ represents synchronous changes across the two series, whereas $\delta=5$ represents changes separated by five time points. Each setting is replicated 500 times. For each replication, we apply the Joint- and Bi-MOSUM procedures using window sizes $G=20$ and $40$. For the CI rule, we use $B=1000$ bootstrap replications.

As the first performance measure, we consider the hit rate, which quantifies the proportion of true change points in $\{Y_t\}$ contained in the estimated hotspots. It is defined as the proportion of trials (out of 500 replications) in which the identified hotspots include the true change points $k^*$ in $\{Y_t\}$. For each replication, in the case of a single change point, a score of 1 is assigned if any identified hotspot contains $k^*$, and 0 otherwise. In the case of two change points, scores of 1, 0.5, or 0 are assigned according to the number of hits.

As the second performance measure, we consider the length of the hotspot interval associated with each true change point. For a true change point contained in $\hat{\mathcal{H}}$, we record the length of the connected hotspot interval containing that change point. If the true change point is not contained in any estimated hotspot, we assign a length of $n=100$, representing the worst case in which the entire observation period would need to be considered. We then average these lengths across true change points and simulation replications. Taken together, the hit rate measures whether the relevant change locations are retained, whereas the interval length measures how precisely the hotspot localizes those changes.

Table \ref{tab:numer2} presents the simulation results. The thresholding rule generally yields higher hit rates than the CI rule, but at the cost of substantially wider hotspot intervals. For the thresholding rule, $G=40$ consistently produces higher hit rates than $G=20$, while also producing considerably longer intervals. For the CI rule, $G=20$ generally yields higher hit rates, particularly when the two series change asynchronously, while the resulting intervals remain substantially shorter than those produced by the thresholding rule. These results illustrate the principal trade-off between the two constructions: the thresholding rule tends to retain true changes over broader temporal regions, whereas the CI rule provides more localized hotspots but may fail to retain some true changes.

The effect of temporal asynchrony differs between the two rules. For the thresholding rule, changing from $\delta=0$ to $\delta=5$ has relatively little effect on the hit rate in both the one- and two-change point settings. This suggests that threshold exceedance regions are sufficiently broad to accommodate a displacement of five time points between changes in the two series. In contrast, the CI rule is more sensitive to the temporal displacement, particularly for $G=40$, for which the hit rate decreases substantially when $\delta$ increases from $0$ to $5$. This behavior is consistent with the more localized nature of the CI-based construction: narrower uncertainty intervals provide greater temporal localization but are less likely to overlap when corresponding changes in the two series occur at different times.

The effect of signal strength on hotspot width also differs between the two constructions. For the thresholding rule, hotspot intervals generally become shorter as the signal weakens because the Mahalanobis distance exceeds the detection threshold over a smaller range of time points. For the CI rule, particularly in the one-change point setting, weaker signals generally lead to greater uncertainty in change point localization and hence wider confidence intervals. This pattern is less systematic when multiple change points are present, where the resulting hotspot lengths also depend on the detection of individual change points and on how their confidence intervals intersect. These contrasting patterns reflect the different mechanisms underlying the two hotspot constructions rather than indicating that one rule is uniformly preferable.

%%%%%%%%%%%%%%%%%%%%%%%%%%%%%%%%%%%%%%%%%%%%%%%%%%%

\begin{table}[t]
\resizebox{\columnwidth}{!}{%
\begin{tabular}{cccccccccccccc}
 &  & \multicolumn{6}{c|}{Hit rate} & \multicolumn{6}{c}{Length of intervals} \\ \hline\hline
\multicolumn{2}{c}{1 change point in $\{Y_t\}$} & Case 1 & Case 2 & Case 3 & Case 4 & Case 5 & Case 6 & Case 1 & Case 2 & Case 3 & Case 4 & Case 5 & Case 6 \\ \hline
\multicolumn{2}{c}{ } & \multicolumn{12}{c}{$\delta=0$} \\ \hline
\multirow{2}{*}{Thrs} 
 & $G=20$ & 0.958 & 0.910 & 0.864 & 0.904 & 0.858 & 0.740 & 28.080 & 25.471 & 23.872 & 25.413 & 21.088 & 18.036 \\ \cline{2-2}
 & $G=40$ & 0.974 & 0.964 & 0.918 & 0.946 & 0.926 & 0.818 & 57.805 & 49.568 & 45.440 & 48.538 & 40.033 & 35.383 \\ \cline{1-2}
 \multirow{2}{*}{CI} 
 & $G=20$ & 0.944 & 0.916 & 0.888 & 0.914 & 0.872 & 0.788 & 1.294 & 2.175 & 2.628 & 2.236 & 4.013 & 6.907 \\ \cline{2-2}
 & $G=40$ & 0.848 & 0.804 & 0.872 & 0.830 & 0.736 & 0.784 & 1.375 & 1.778 &   2.515 & 1.901 & 3.743 & 6.859 \\ \hline
\multicolumn{2}{c}{ } & \multicolumn{12}{c}{$\delta=5$} \\ \hline
\multirow{2}{*}{Thrs} 
  & $G=20$ & 0.956 & 0.902 & 0.876 & 0.912 & 0.842 & 0.728 & 28.431 & 25.114 & 24.685 & 24.550 & 20.160 & 17.844 \\ \cline{2-2}
 & $G=40$ & 0.976 & 0.936 & 0.918 & 0.944 & 0.916 & 0.812 & 59.492 & 51.872 & 50.051 & 50.126 & 40.966 & 36.963 \\ \cline{1-2}
 \multirow{2}{*}{CI} 
 & $G=20$ & 0.736 & 0.628 & 0.594 & 0.556 & 0.504 & 0.494 & 3.036 & 3.227 & 2.888 & 2.601 & 4.138 & 5.946 \\ \cline{2-2}
 & $G=40$ & 0.298 & 0.304 & 0.270 & 0.264 & 0.290 & 0.348 & 1.337 & 1.767 & 2.000 & 1.706 & 3.606 & 6.311 \\ \hline\hline 
\multicolumn{2}{c}{2 change points in $\{Y_t\}$} & Case 1 & Case 2 & Case 3 & Case 4 & Case 5 & Case 6 & Case 1 & Case 2 & Case 3 & Case 4 & Case 5 & Case 6 \\ \hline
\multicolumn{2}{c}{ } & \multicolumn{12}{c}{$\delta=0$} \\ \hline
\multirow{2}{*}{Thrs} 
 & $G=20$ & 0.797 & 0.736 & 0.658 & 0.702 & 0.589 & 0.525 & 45.346 & 40.866 & 37.877 & 40.328 & 32.620 & 28.086 \\ \cline{2-2}
 & $G=40$ & 0.956 & 0.937 & 0.902 & 0.910 & 0.836 & 0.780 & 63.868 & 58.092 & 54.345 & 55.388 & 45.745 & 39.743\\ \cline{1-2}
\multirow{2}{*}{CI} 
 & $G=20$ & 0.672 & 0.628 & 0.578 & 0.620 & 0.574 & 0.490 & 18.051 & 19.379 & 16.808 & 18.340 & 20.309 & 16.533\\ \cline{2-2}
 & $G=40$ & 0.491 & 0.425 & 0.382 & 0.429 & 0.365 & 0.351 & 15.291 & 12.990 & 14.600 & 14.094 & 12.472 & 15.159\\ \hline
\multicolumn{2}{c}{ } & \multicolumn{12}{c}{$\delta=5$} \\ \hline
\multirow{2}{*}{Thrs} 
 & $G=20$ & 0.792 & 0.725 & 0.654 & 0.674 & 0.582 & 0.521 & 45.622 & 41.228 & 36.310 & 38.758 & 32.332 & 26.940 \\ \cline{2-2}
 & $G=40$ & 0.961 & 0.946 & 0.900 & 0.904 & 0.842 & 0.788 & 63.374 & 58.078 & 52.133 & 54.952 & 45.037 & 39.676\\ \cline{1-2}
\multirow{2}{*}{CI} 
 & $G=20$ & 0.728 & 0.629 & 0.508 & 0.572 & 0.493 & 0.410 & 24.570 & 22.801 & 17.624 & 19.679 & 19.249 & 15.591 \\ \cline{2-2}
 & $G=40$ & 0.214 & 0.195 & 0.181 & 0.187 & 0.218 & 0.194 & 7.994 & 7.208 & 7.494 & 7.688 & 10.026 & 9.072 \\ \hline
\end{tabular}%
}
\caption{Average hit rates and hotspot interval lengths for the thresholding (Thrs) and confidence-interval (CI) rules. The hit rate is the proportion of true change points in $\{Y_t\}$ contained in the estimated hotspots. For each true change point, interval length is the length of the connected hotspot interval containing that change point; a value of $n=100$ is assigned when the change point is not contained in an estimated hotspot. The significance level is $\alpha=0.05$.}
\label{tab:numer2}
\end{table}

%%%%%%%%%%%%%%%%%%%%%%%%%%%%%%%%%%%%%%%%%%%%%%%%%%%

\subsection{Data Application}\label{sec:data_appl}

We apply the hotspot rules examined in Section \ref{ssec:hotspots_simulation} to data from the ALACRITY Phase I study, conducted by the Weill Cornell ALACRITY Center (P50MH113838), involving middle-aged and older adults with depression undergoing psychotherapy. For illustration, we consider four patients (sternrelief15, sternrelief16, sternrelief5, and sternrelief6) and examine their daily emotional stress levels together with three passive sensing variables: step counts, conversation percentage, and sleep duration. These variables represent different behavioral domains, namely physical activity, sociability, and sleep behavior, which have been associated with stress \cite[e.g.,][]{triantafillou2019relationship,dasilva2021daily,baretta2025predicting}. The passive sensing measurements were collected through an application installed on the patients' smartphones. Patients reported their stress levels daily using a 5-point Likert scale (1 = none, 5 = extremely high), and the passive sensing variables were summarized daily to match the observation frequency of the stress measurements. Because the passive sensing variables are measured on different scales, we analyze each variable in its original units without normalization.

Since the true change locations are unknown in this real-data application, our objective is not to assess detection accuracy but to illustrate how the proposed procedures summarize temporally related changes in stress and passive sensing variables. We present sternrelief5 with $G=40$ as a representative example in Figure \ref{fig:hostpot_sternrelief5}. For completeness, Table \ref{tab:data_appl} in Appendix \ref{ap:appendix} reports the estimated change points and identified hotspots for all four patients, passive sensing variables, and window sizes. The corresponding visualizations for sternrelief15, sternrelief16, and sternrelief6 are also provided in Appendix \ref{ap:appendix}.

\begin{figure}[t]
\centering
\includegraphics[width=\textwidth,height=0.5\textwidth]{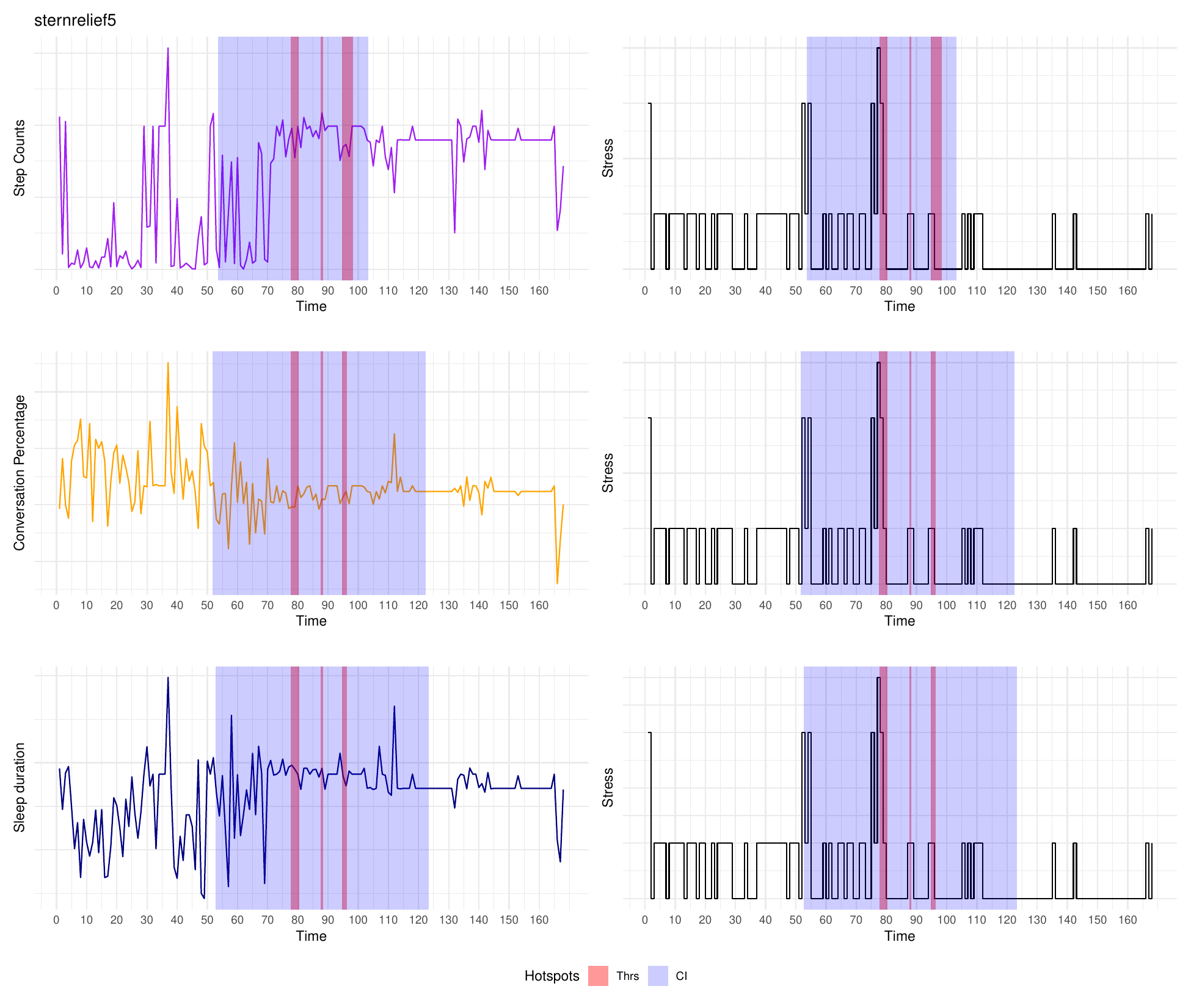}
\caption{Results of the identified hotspots for sternrelief5 with $G=40$ across three combinations of passive sensing variables (left; step count, conversation percentage, and sleep duration in each row) and stress levels (right). The two definitions, threshold and confidence interval rules, are shown as red and blue shaded areas, respectively.}
\label{fig:hostpot_sternrelief5}
\end{figure}

Figure \ref{fig:hostpot_sternrelief5} illustrates the motivation for representing detected changes as hotspot intervals rather than as isolated change point estimates. Changes in the passive sensing variables and stress levels do not necessarily occur simultaneously, but the proposed procedures can summarize changes occurring within nearby temporal regions. The thresholding rule identifies relatively localized regions in which the relevant Mahalanobis distances exceed the detection threshold, whereas the confidence-interval rule incorporates uncertainty in the estimated change locations and consequently produces a broader temporal representation. This distinction is visible across all three passive sensing variables in Figure \ref{fig:hostpot_sternrelief5}, illustrating how the two rules provide complementary summaries of potentially asynchronous changes.

The complete results in Table \ref{tab:data_appl} and the additional figures in Appendix \ref{ap:appendix} further show that the identified hotspots can vary with both the passive sensing variable and the choice of window size $G$. Nevertheless, similar temporal regions are identified across step counts, conversation percentage, and sleep duration in several instances, suggesting that the identified patterns are not necessarily driven by a single behavioral measure. Differences between $G=20$ and $G=40$ are expected because the window size determines the temporal scale over which distributional changes are evaluated. We therefore regard hotspot regions that recur across multiple passive sensing variables or window sizes as more stable across analysis configurations, without interpreting any individual detected change point as a known ground-truth event.

%%%%%%%%%%%%%%%%%%%%%%%%%%%%%%%%%%%%%%%%%%%%%%%%%%%

\section{Discussion}\label{sec:discuss}

In this article, we develop a framework for identifying hotspots, defined as temporal regions in which changes in stress and passive sensing variables occur at the same or nearby times. The proposed approach is intended as an auxiliary tool for examining temporal relationships between stress and behavioral measures derived from passive sensing data. Building on MOSUM-type CPD methods, the Joint- and Bi-MOSUM procedures detect changes in statistical features within and across series, and the resulting change point information is combined through thresholding and confidence-interval rules to construct hotspot regions. In this way, the framework summarizes potentially asynchronous change point detections as interpretable temporal regions rather than isolated time points.

The simulation studies examined both change point detection and hotspot construction. When changes in multiple distributional features coexist, the Joint- and Bi-MOSUM procedures generally achieve higher detection power and better control of false detections than methods targeting individual changes. The hotspot simulations further demonstrate a trade-off between the two proposed rules: the thresholding rule tends to retain true changes over broader temporal regions, whereas the confidence-interval rule provides more localized regions but can be more sensitive to temporal asynchrony. In the real-data application, hotspot regions occurring at similar times are observed across multiple passive sensing variables, illustrating how the framework can reveal temporal concordance between changes in stress and different aspects of behavior without requiring the changes to occur simultaneously.

Several limitations suggest directions for future work. First, the current framework analyzes passive sensing variables separately rather than jointly in a fully multivariate model. Extending the Mahalanobis-distance-based construction to accommodate multiple passive sensing variables and multiple distributional features simultaneously would require additional methodological and theoretical development. Second, when different passive sensing variables yield different hotspot regions, methods for integrating or prioritizing these results may be useful. Finally, the proposed framework identifies temporal concordance but does not formally establish an association, and should not be interpreted as providing evidence of causality between passive sensing variables and stress. Future work could integrate the hotspot framework with statistical models that formally evaluate these relationships in clinical studies.

%%%%%%%%%%%%%%%%%%%%%%%%%%%%%%%%%%%%%%%%%%%%%%%%%%%
\section{Software}

The R code used for illustrating the method in Section \ref{ssec:illustration} and for the simulation studies and data application in Section \ref{sec:numerical} is available on \texttt{GitHub} at \href{https://github.com/yk748/hotspot}{https://github.com/yk748/hotspot}.

%%%%%%%%%%%%%%%%%%%%%%%%%%%%%%%%%%%%%%%%%%%%%%%%%%%
\section*{Acknowledgements}

The data that support the findings of this study are available on request from the corresponding author. The data are not publicly available due to privacy or ethical restrictions.

\section*{Funding}

S. Banerjee acknowledges partial supports from NIMH award P50MH113838 and NIA award P01AG073090. S. Basu acknowledges partial support from NSF CAREER award DMS-2239102. In addition, S. Basu acknowledges partial support from NSF awards DMS-1812128, DMS-2210675, and NIH awards R01GM135926, R21NS120227.

%%%%%%%%%%%%%%%%%%%%%%%%%%%%%%%%%%%%%%%%%%%%%%%%%%%

\appendix

\setcounter{section}{0}
\setcounter{table}{0}
\setcounter{figure}{0}
\setcounter{equation}{0}
\renewcommand{\thesection}{A.\arabic{section}}
\renewcommand{\thetable}{A.\arabic{table}}
\renewcommand{\thefigure}{A.\arabic{figure}}
\renewcommand{\theequation}{A.\arabic{equation}}

\section{Additional Tables and Figures}\label{ap:appendix}

%%%%%%%%%%%%%%%%%%%%%%%%%%%%%%%%%%%%%%%%%%%%%%%%%%%

\begin{table}[h]
\resizebox{\columnwidth}{!}{%
\begin{tabular}{cccccccccc}
 &  &  & \multicolumn{5}{c}{Change points} & \multicolumn{2}{|c}{Hotspots} \\ \hline
Patients & $G$ & Variables & $Y$ & $YX$ & $YX^2$ & $Y^2X$ & $Y^2X^2$ & Thrs & CI \\ \hline
\multirow{6}{*}{sternreleif15} 
& \multirow{3}{*}{$G=20$} 
& SC & - & - & - & 96 & - & - & -  \\ \cline{3-3}
 &  & CP & - & - & - & - & - & - & - \\ \cline{3-3}
 &  & SD & - & 62 & - & 62 & - & - & -  \\ \cline{2-3}
 & \multirow{3}{*}{$G=40$} 
 & SC & 59,81 & 63,82,109 & 82,93 & 101 &  & [59,59],[81,83],[93,93] & [30,109] \\ \cline{3-3}
 &  & CP & 59,81 & 81 & 63,82,93 & - & 62 & [57,57],[59,59],[81,88],[91,93] & [38,107] \\ \cline{3-3}
 &  & SD & 59,81 & 62,82,101,126 & 82,93 & 62,101,126 & [57,57],[59,59],[81,82],[91,93] & [30,107]  \\ \hline
\multirow{6}{*}{sternreleif16} 
& \multirow{3}{*}{$G=20$} 
& SC & 44,64 & 44,64 & 45,64 & 45,64 & - & [43,47],[63,65],[87,87] & [24,83]  \\ \cline{3-3}
 &  & CP & 44,64 & 45,64 & 45,65 & 36,56 & 36,50,59 & [43,52],[62,65],[87,87] & [24,83] \\ \cline{3-3}
 &  & SD & 44,64 & 30,45,64 & 44,64 & 30,51 & 56 & [43,52],[63,64],[87,87] & [24,83] \\ \cline{2-3}
 & \multirow{3}{*}{$G=40$} 
 & SC & 44 & - & - & - & - & - & -  \\ \cline{3-3}
 &  & CP & 44 & - & 45 & - & - & [42,47],[87,87] & [10.98,77.03] \\ \cline{3-3}
 &  & SD & 44 & - & 44 & - & - & [44,45],[87,87] & [9.98,78.03] \\ \hline
\multirow{6}{*}{sternreleif5} 
& \multirow{3}{*}{$G=20$} 
& SC & - & 72,80 & 78,131 & 72,93,100 & 78 & - & - \\ \cline{3-3}
 &  & CP & - & - & 80,132 & - & 70 & - & - \\ \cline{3-3}
 &  & SD & - & 112,132 & - & - & 69 & - & - \\ \cline{2-3}
 & \multirow{3}{*}{$G=40$} 
 & SC & 79,96 & 50,77 & 83 & 50,77 & 83 & [78,80],[88,88],[95,98] & [54,103] \\ \cline{3-3}
 &  & CP & 79,96 & 52,79,91 & 80 & 50 & 72 & [78,80],[88,88],[95,96] & [52.00,122.03] \\ \cline{3-3}
 &  & SD & 79,96 & 49,78,112 & 78 & 49,72,112 & 72 & [78,80],[88,88],[95,96] & [53,123] \\ \hline
\multirow{6}{*}{sternreleif6} 
& \multirow{3}{*}{$G=20$} 
& SC & 62 & 62 & 62 & - & - & [53,70] & [52,72] \\ \cline{3-3}
 &  & CP & 62 & 62 & 62,89 & 66 & 89 & [53,70] & [52,72] \\ \cline{3-3}
 &  & SD & 62 & 62 & 64 & - & - & [53,70] & [53,71] \\ \cline{2-3}
 & \multirow{3}{*}{$G=40$} 
 & SC & 54,63 & 54,63 & 9,54,63 & - & 9 & [43,79] & [36.98,80.00] \\ \cline{3-3}
 &  & CP & 54,63 & 54,63 & 54,63 & 69 & 70 & [43,79] & [35,82] \\ \cline{3-3}
 &  & SD & 54,63 & 54,63 & 54,63 & - & - & [44,79] & [32.00,81.03] \\ \hline
\end{tabular}%
}
\caption{Results of the estimated change points (middle five columns) are shown for different types of changes, along with the identified hotspots (right two columns). Each of the four sections represents one patient, with three passive sensing variables considered per patient and for each window size, $G=20,40$. The abbreviations SC, CP, and SD in the variable names denote step counts, conversation percentage, and sleep duration, respectively. A significance level of $\alpha=0.05$ is used for all steps.}
\label{tab:data_appl}
\end{table}

%%%%%%%%%%%%%%%%%%%%%%%%%%%%%%%%%%%%%%%%%%%%%%%%%%%

\begin{figure}[h]
\centering
\includegraphics[width=1\textwidth,height=0.35\textheight]
{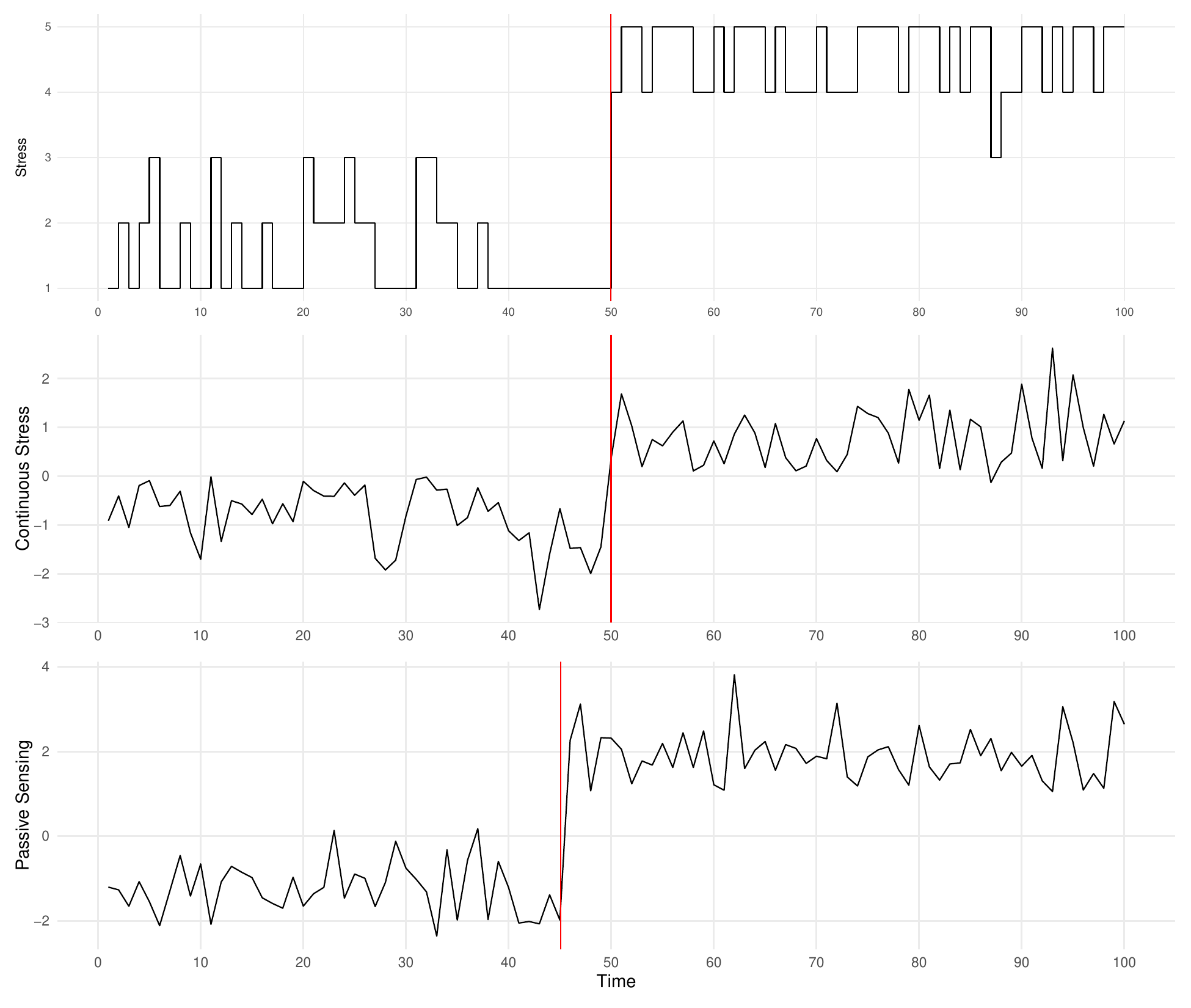}
\includegraphics[width=1\textwidth,height=0.35\textheight]
{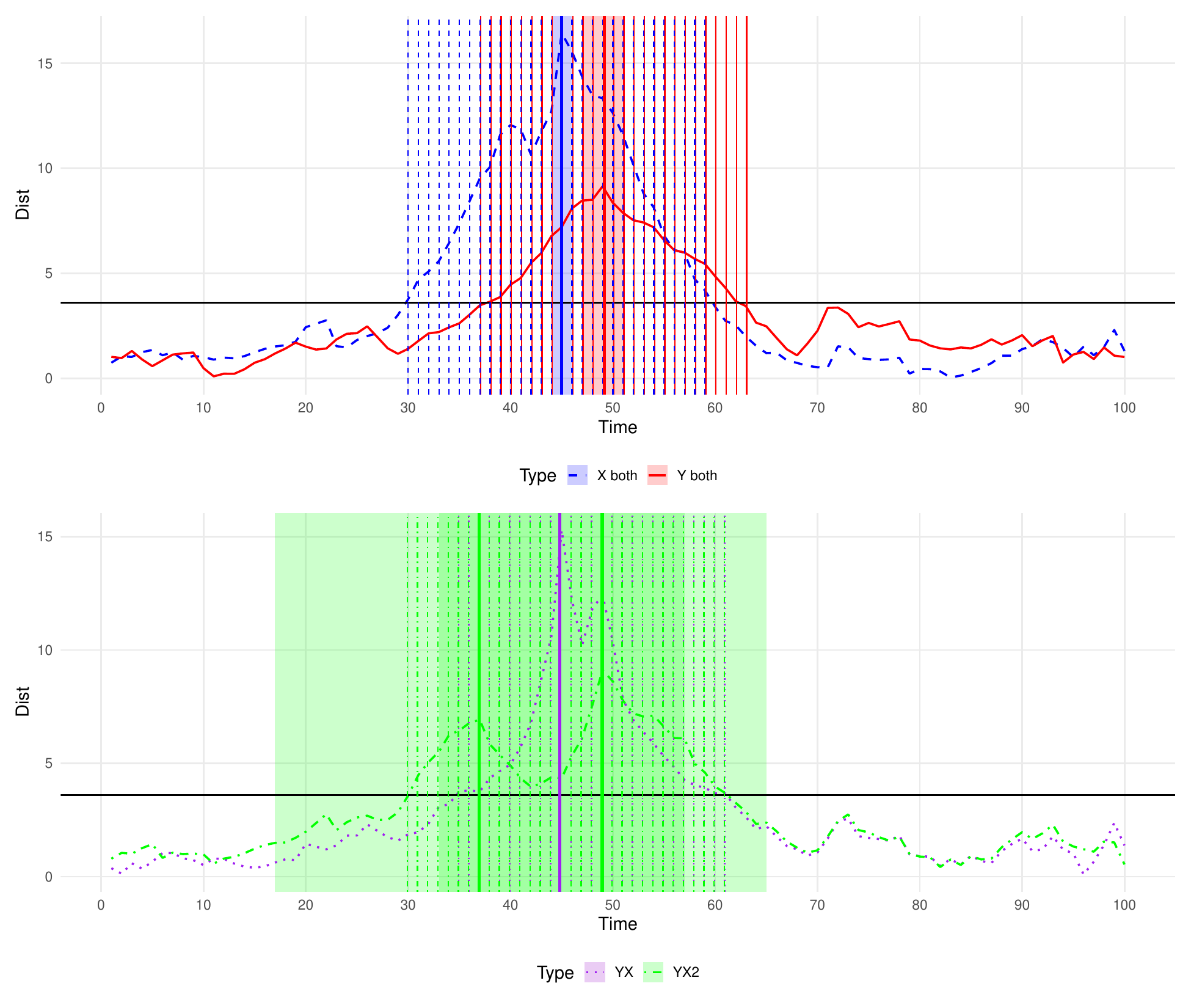}
\caption{(Top) Time series plots of stress levels (first), their continuous counterparts ($\{Y_t\}$; second), and the passive sensing variable ($\{X_t\}$; third). The true change point of the stress level is at $k^*=50$. (Bottom) Mahalanobis distances from the Joint-MOSUM of $Y_t$ and $X_t$ are shown in red and blue (first), and those from the Bi-MOSUM for mean changes in both $Y_t$ and $X_t$ (purple) and for a mean change in $Y_t$ with a variance change in $X_t$ (green). The thick vertical lines indicate the estimated change points from each MOSUM. The dashed vertical lines indicate the time points at which each distance exceeds the threshold (black), which is externally computed using a Monte Carlo simulation (slightly different from the one used in the MOSUM algorithm). The shaded areas represent pointwise confidence intervals. The hotspots are thus $[37,61]$ by the thresholding rule and $[47,51]$ by the confidence interval rule, respectively.}
 \label{fig:hotspot_mean}
\end{figure}

\begin{figure}[h]
\centering
\includegraphics[width=1\textwidth,height=0.35\textheight]
{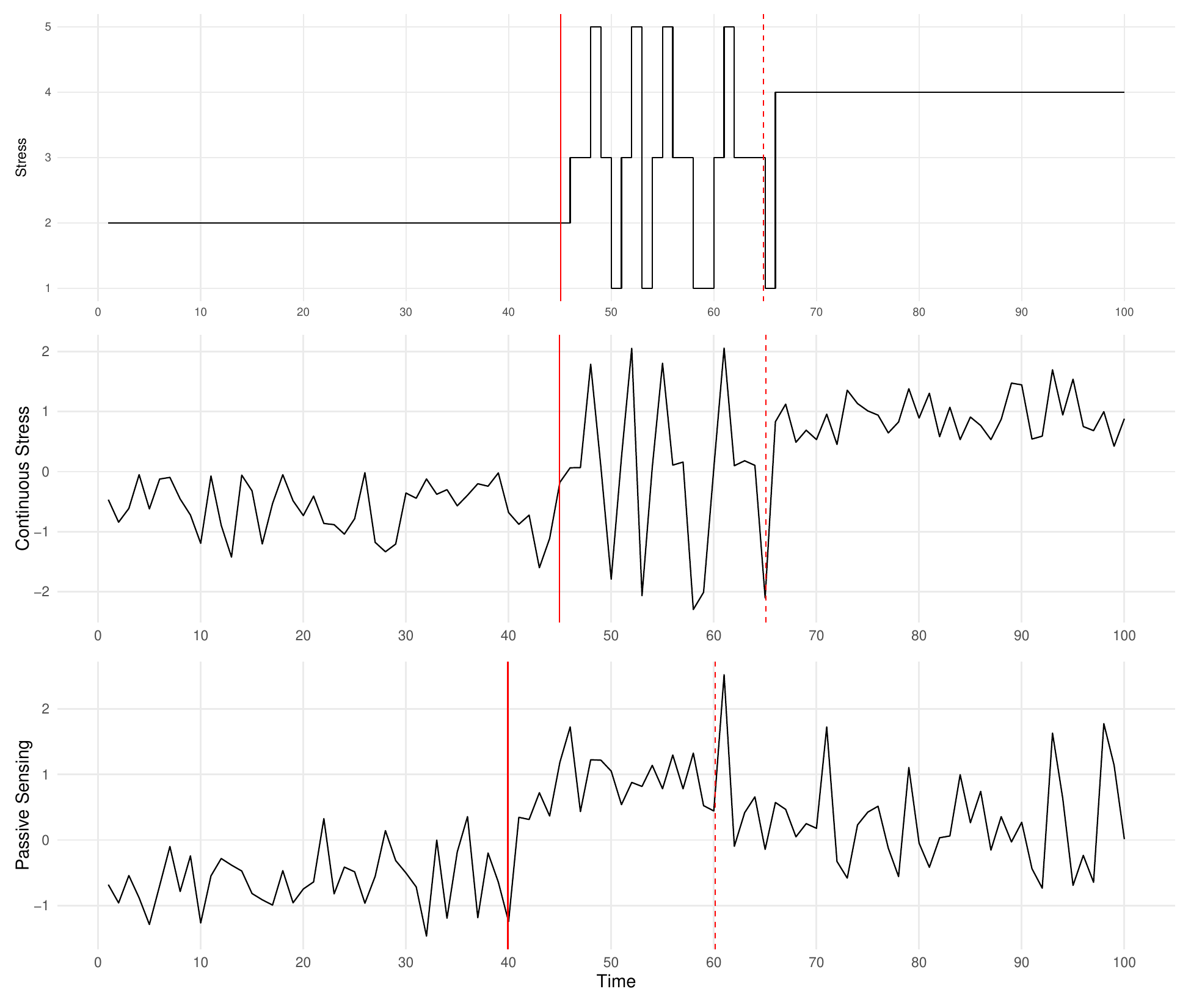}
\includegraphics[width=1\textwidth,height=0.35\textheight]
{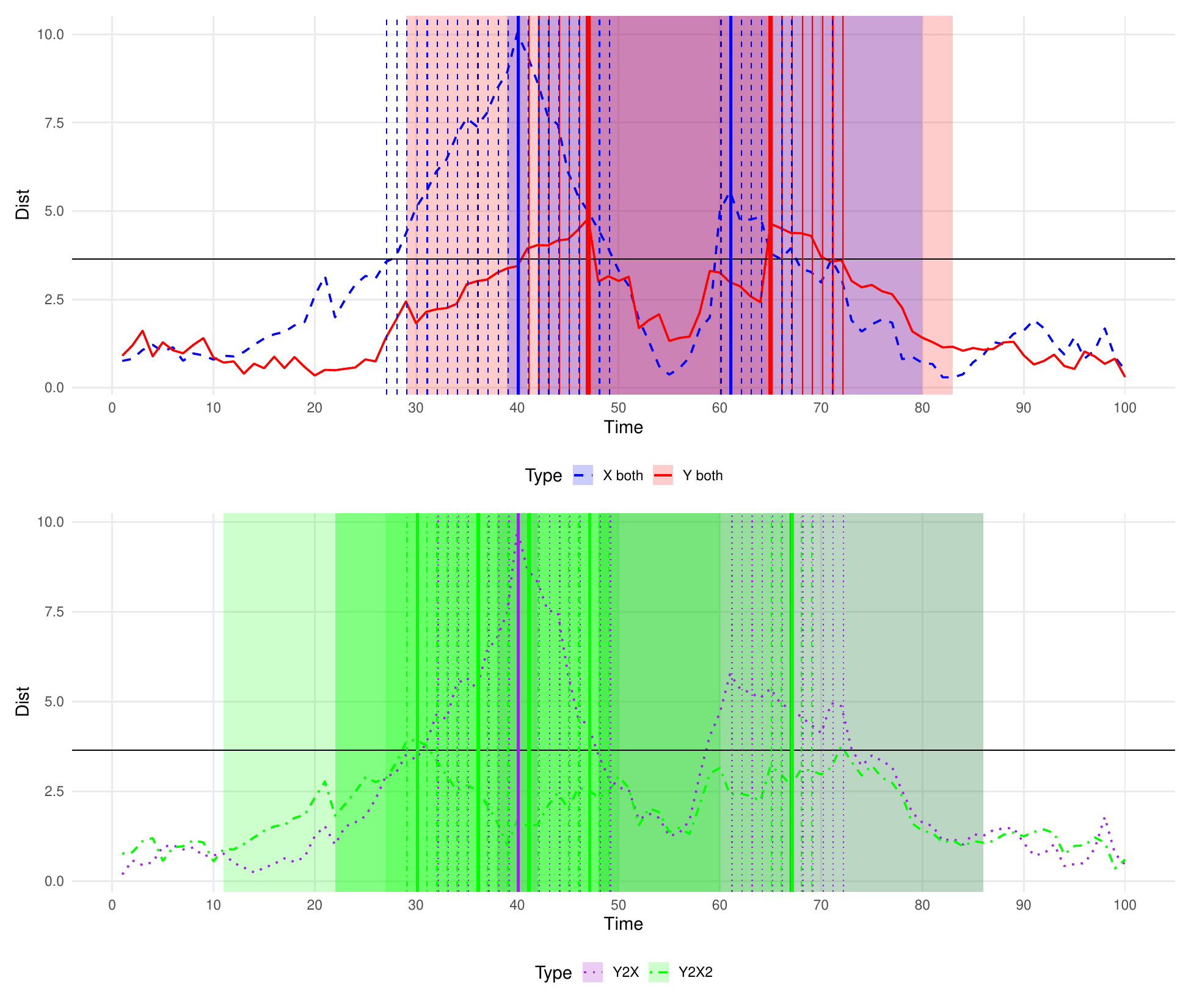}
\caption{(Top) Time series plots of stress levels (first), their continuous counterparts ($\{Y_t\}$; second), and the passive sensing variable ($\{X_t\}$; third). The true change points of the stress level are at $k^*=40,60$. (Bottom) Mahalanobis distances from the Joint-MOSUM of $Y_t$ and $X_t$ are shown in red and blue (first), and those from the Bi-MOSUM for variance change in $Y_t$ with a mean change in $X_t$ (purple) and both variance changes in both $Y_t$ and $X_t$ (green). The thick vertical lines indicate the estimated change points from each MOSUM. The dashed vertical lines indicate the time points at which each distance exceeds the threshold (black), which is externally computed using a Monte Carlo simulation (slightly different from the one used in the MOSUM algorithm). The shaded areas represent uniform confidence intervals. The hotspots are thus $[40,47],[65,72]$ by the thresholding rule and $[29,83]$ by the confidence interval rule, respectively.}
 \label{fig:hotspot_var}
\end{figure}

%%%%%%%%%%%%%%%%%%%%%%%%%%%%%%%%%%%%%%%%%%%%%%%%%%%

\begin{figure}[h]
\centering
\includegraphics[width=\textwidth,height=0.5\textwidth]{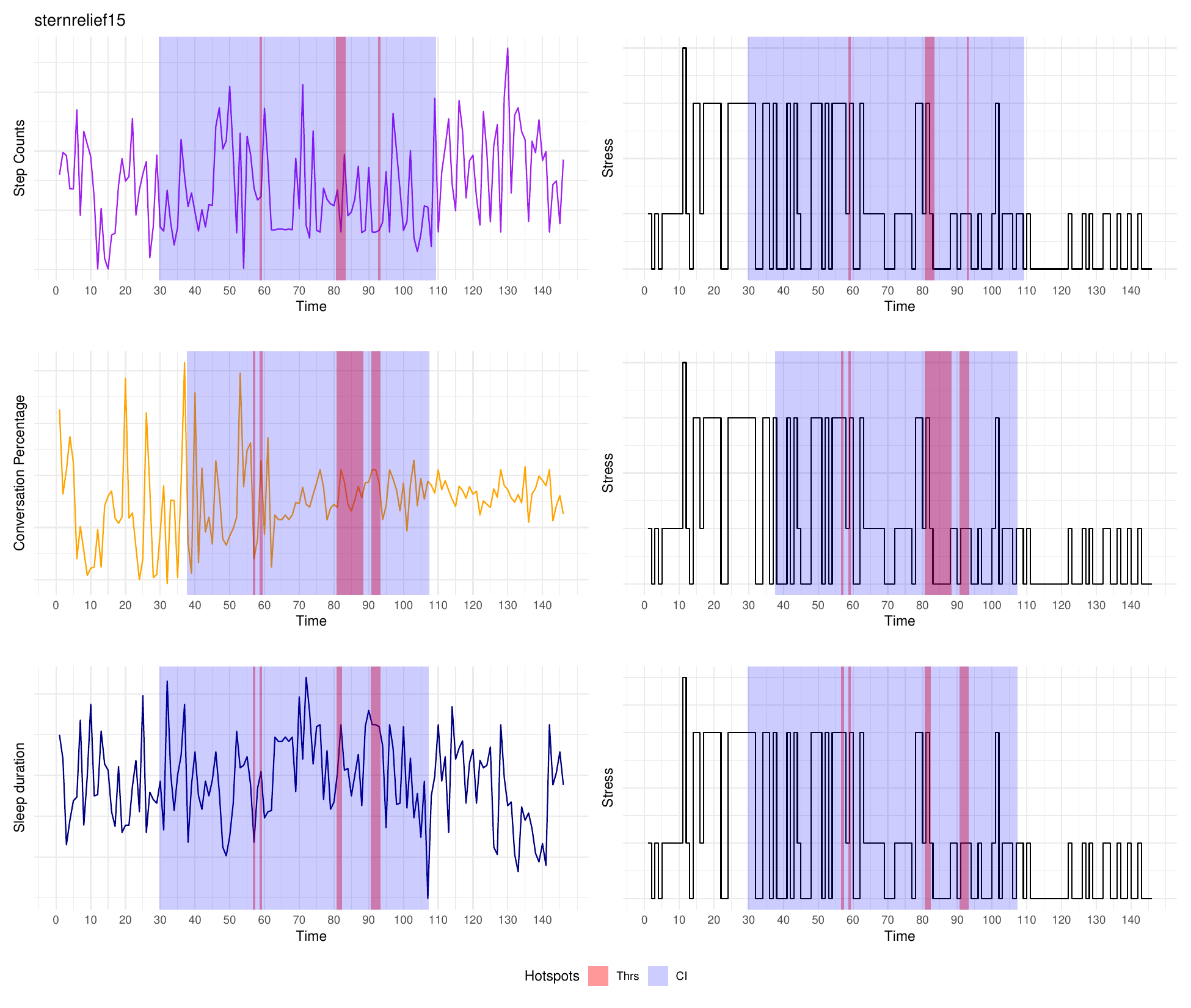}
\caption{Results of the identified hotspots for sternrelief15 with $G=40$ across three combinations of passive sensing variables (left; step count, conversation percentage, and sleep duration in each row) and stress levels (right). The two definitions, threshold and confidence interval rules, are shown as red and blue shaded areas, respectively.}
\label{fig:hostpot_sternrelief15}
\end{figure}

%%%%%%%%%%%%%%%%%%%%%%%%%%%%%%%%%%%%%%%%%%%%%%%%%%%

\begin{figure}[b]
\centering
\includegraphics[width=\textwidth,height=0.5\textwidth]{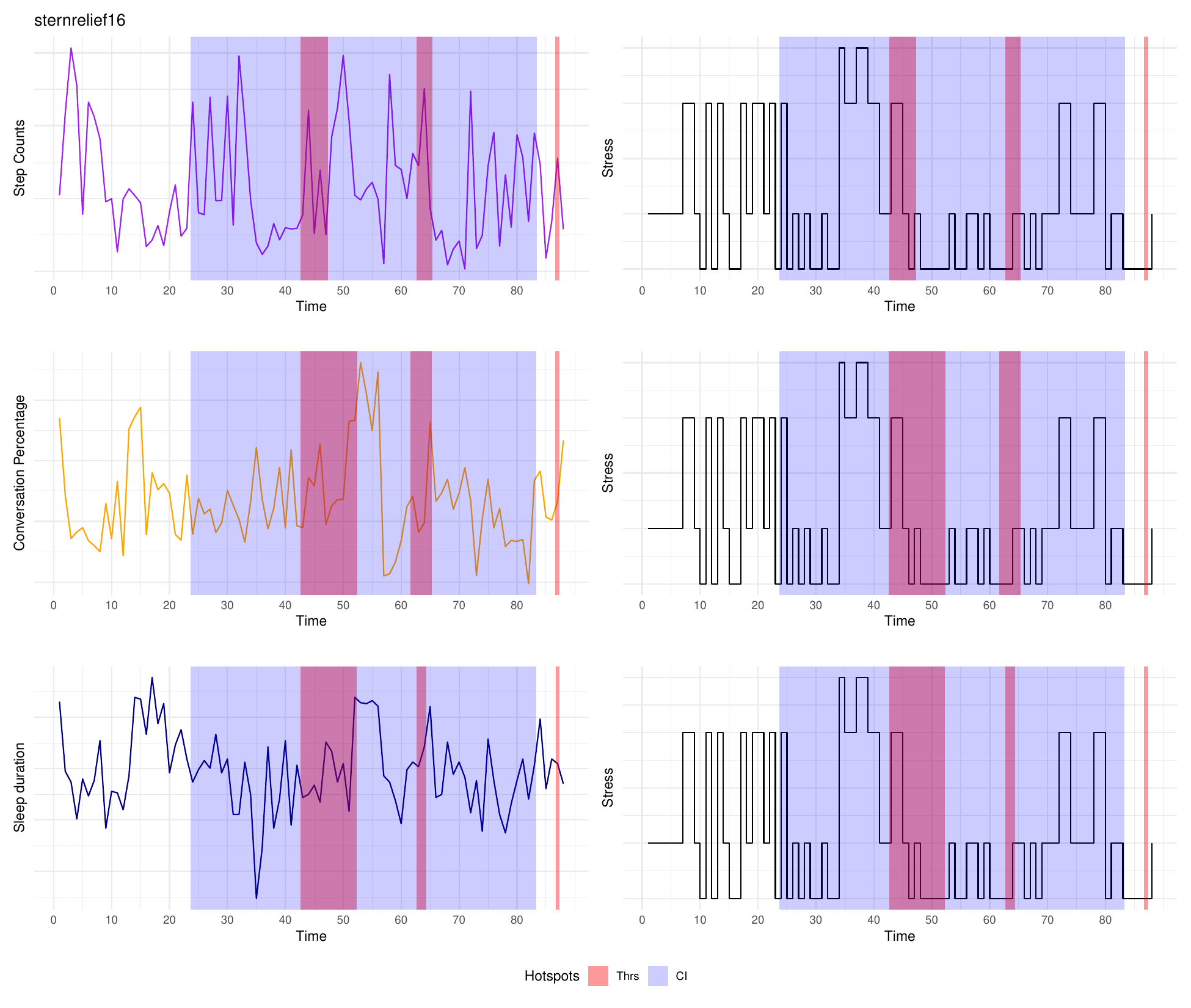}
\caption{Results of the identified hotspots for sternrelief16 with $G=20$ across three combinations of passive sensing variables (left; step count, conversation percentage, and sleep duration in each row) and stress levels (right). The two definitions, threshold and confidence interval rules, are shown as red and blue shaded areas, respectively.}
\label{fig:hostpot_sternrelief16}
\end{figure}

%%%%%%%%%%%%%%%%%%%%%%%%%%%%%%%%%%%%%%%%%%%%%%%%%%%

\begin{figure}[t]
\centering
\includegraphics[width=\textwidth,height=0.5\textwidth]{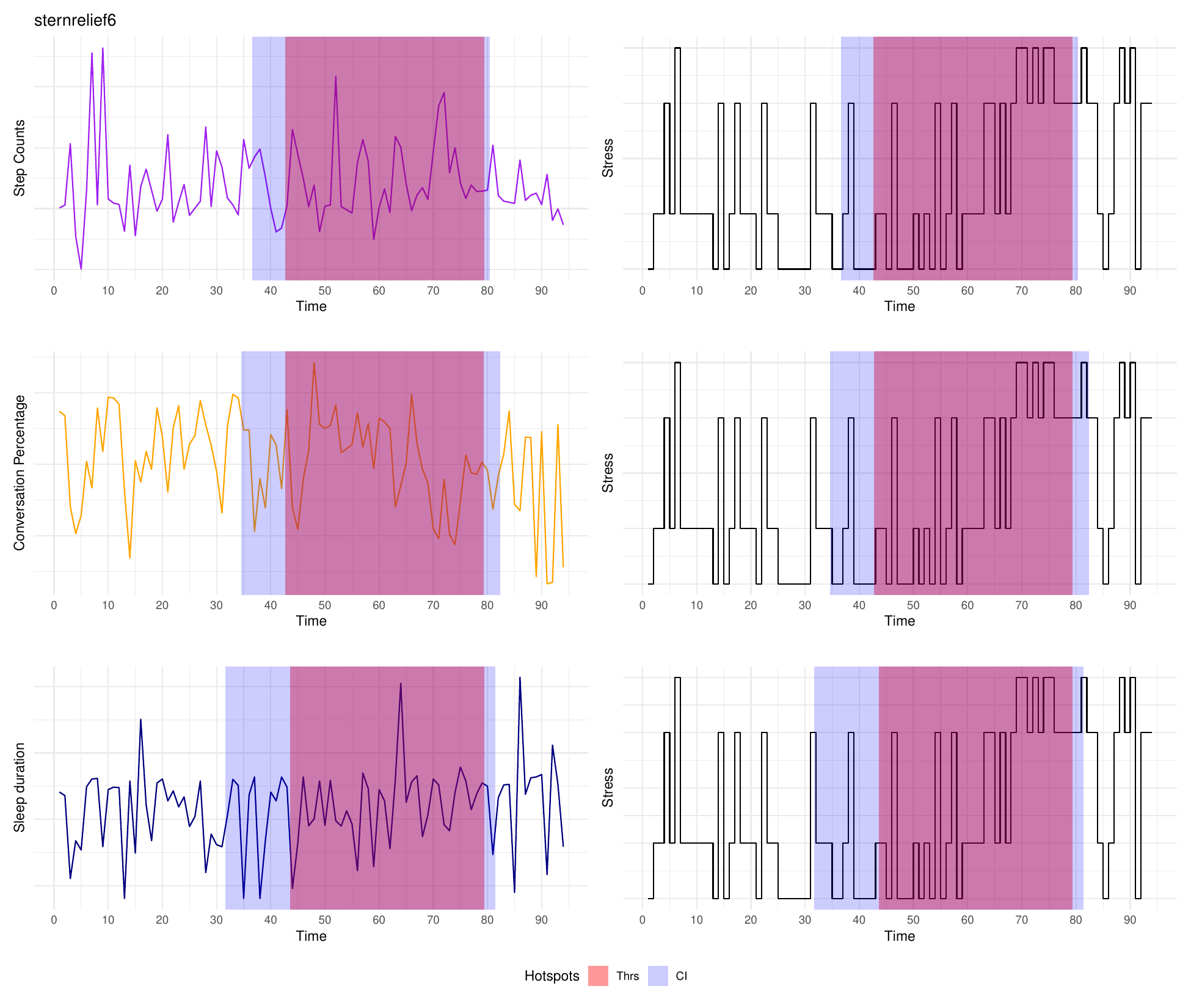}
\caption{Results of the identified hotspots for sternrelief6 with $G=40$ across three combinations of passive sensing variables (left; step count, conversation percentage, and sleep duration in each row) and stress levels (right). The two definitions, threshold and confidence interval rules, are shown as red and blue shaded areas, respectively.}
\label{fig:hostpot_sternrelief6}
\end{figure}

%%%%%%%%%%%%%%%%%%%%%%%%%%%%%%%%%%%%%%%%%%%%%%%%%%%

\FloatBarrier

\bibliographystyle{apalike}
\bibliography{hotspot}

\end{document}